\documentclass[compress]{elsarticle}
\usepackage{lineno,hyperref}
\modulolinenumbers[10]
\usepackage{geometry}
\usepackage{color}
\usepackage{subfigure}
\usepackage{float}
\usepackage{amsfonts}
\usepackage{amsmath}
\usepackage{color}
\usepackage{caption}
\usepackage{bm}
\usepackage{algorithm}
\usepackage{algorithmic}
\usepackage{multirow}
\usepackage{amssymb}
\usepackage{array}
\usepackage{mathrsfs}
\usepackage{upgreek}

\usepackage{makecell}
\usepackage{graphicx}

\journal{Arxiv}

\begin{document}

	\begin{frontmatter}
		\title{Multi-time-step coupling of peridynamics and classical continuum mechanics for dynamic brittle fracture}
		
		\author[a]{Jiandong Zhong}
		\author[a,b]{Fei Han\corref{mycorrespondingauthor}}
		\cortext[mycorrespondingauthor]{Corresponding author}
		\ead{hanfei@dlut.edu.cn}
		
		\author[a,b]{Zongliang Du}
		\author[a,b]{Xu Guo\corref{mycorrespondingauthor}}
		\ead{guoxu@dlut.edu.cn}
	
		\address[a]{State Key Laboratory of Structural Analysis, Optimization and CAE Software for Industrial Equipment, Department of Engineering Mechanics, Dalian University of Technology, Dalian 116023, China.}
		\address[b]{Ningbo Institute of Dalian University of Technology, Dalian University of Technology, Ningbo 315016, China}
		
		\begin{abstract}	
		Peridynamics (PD), as a nonlocal theory, is well-suited for solving problems with discontinuities, such as cracks. However, the nonlocal effect of peridynamics makes it computationally expensive for dynamic fracture problems in large-scale engineering applications. As an alternative, this study proposes a multi-time-step (MTS) coupling model of PD and classical continuum mechanics (CCM) based on the Arlequin framework. Peridynamics is applied to the fracture domain of the structure, while continuum mechanics is applied to the rest of the structure. The MTS method enables the peridynamic model to be solved at a small time step and the continuum mechanical model is solved at a larger time step. Consequently, higher computational efficiency is achieved for the fracture domain of the structure while ensuring computational accuracy, and this coupling method can be easily applied to large-scale engineering fracture problems.
		\end{abstract}
	\end{frontmatter}

	\section{Introduction}
	The failure mechanisms of structures under dynamic loading have always been a difficult problem in the field of engineering science. Dynamic fracture is more complex and more common in our daily lives compared to fractures caused by quasi-static loading. It plays a significant role in various fields such as energy engineering, materials science, topological optimization of structures, and so on. It is crucial to understand the dynamic behavior of fractures for the prediction and prevention of catastrophic failures in structures, as well as the design of materials with enhanced fracture resistance. 

	The dynamic fracture problem is the fundamental in the science of fracture. Initially, a wide range of phenomena has been observed in experiments on dynamic brittle materials, such as crack branching and crack path instability \cite{Ramulu1983,Ramulu1985}. 
	It is a challenge to model failure or fracture problems by using classical continuum mechanics (CCM), because of its dependence on partial differential equations that are undefined along discontinuities in mathematical form. Over the past several decades with the development of computer science and computational mechanics, significant efforts have been dedicated to simulating phenomena of dynamic brittle fracture, which result in a notable progress of dynamic fracture in the CCM framework. However, many numerical methods for simulating cracks may contradict basic partial differential equations in CCM, such as meshfree techniques and extended finite element methods (XFEM) \cite{XfemBely,EFGBely,BelytschkoEFG}. These methods require adding supplementary conditions to approximate discontinuous displacement fields for predicting crack growth, which complicates model processing and affects the efficiency and accuracy of the simulations. Thus, predicting dynamic fracture in brittle materials is still an open problem as simulations often fail to reproduce many experimentally observed features of dynamic brittle fracture. The ability to accurately predict and control fracture behavior under dynamic loading conditions remains a fascinating and challenging pursuit, with wide-ranging implications for engineering science.

	In order to overcome the drawback in CCM, Silling \cite{SILLING2000175} proposed a nonlocal mechanics theory called peridynamics (PD) that describes material motion using an integro-differential equation. At first, the bond-based PD assumes that one material point can interact with others in its neighborhood using bonds, which can be seen as springs \cite{SILLING20051526}. Under this assumption, the PD theory no longer requires the continuous displacement field in CCM. The main advantage of PD is that it can spontaneously predict damage and crack growth in a structure without using additional technologies because the governing equations in PD remain valid even if discontinuities appear in the material.  Although the theory is still in its infancy, the literature on PD  has been fairly rich and extensive in the last few decades \cite{PDreview,madenci2022advances}. Consequently, there is a new perspective for modeling dynamic fracture problems with the help of PD. In bond-based PD, Silling \cite{SILLING2003641} successfully verified this theory by modeling a classical dynamic fracture problem; Ha and Bobaru \cite{Bobaru2010,HA20111156} studied the bifurcation problem of cracks and the characteristics of dynamic brittle fracture; Bobaru and Hu \cite{Bobaru2012} discussed the influence of the peridynamic horizon on crack branching in brittle materials; then on the topic of why cracks branch, Bobaru and Zhang \cite{Bobaru2015} carried out systematic analysis and research; Yu \cite{Yu2023,FAN2022115340,TRASK2019151} presented a reliable computational approach for describing material heterogeneity and brittle fractures; and a few bond-based PD models were proposed for analyzing dynamic crack propagation in different problems such as orthotropic media \cite{GHAJARI2014431}, functionally graded materials \cite{CHENG2015529,cmes.2019.06374}, and perforated plates \cite{Holechen}. Besides, Galvanetto and co-workers \cite{Ugo2014} proposed an adaptive grid refinement method in two-dimensional PD; Imachi \cite{IMACHI2019359} enhanced state-based PD by using a transition bond model which suppressed the numerical oscillation in the model for dynamic fracture analysis. Furthermore, practical applications of PD have also been developed, such as wave isolation \cite{SAJAL2023108456}, elastic instability and failure in lattice beam structures \cite{ROY2023116210}, and shape design optimization \cite{OH2021107837}.

	However, one of the primary concerns is the increased computational cost due to the nonlocality of PD. This makes computations expensive, thus posing limitations in engineering applications. Presently, there are two main approaches to solving this problem: one approach is to use high-performance computing (HPC) to accelerate the computation process, this can indeed tackle some issues in a few studies \cite{CUDAPD,MOSSAIBY20171856,WANG2022103458,Peridigm}, but with the increasing scale of computation, the challenge of using PD still exists because HPC only accelerates the computation speed without reducing the computational cost; and another approach is the coupling method in which PD is only applied in the fracture domain and the rest is modeled by the local model like CCM to save computational cost \cite{YU2021113962,LIU2012163,HAN2016336,ZACCARIOTTO2018471,Lee2016ParallelPO,KEFAL2022114520,Anicode2023,YU2018905,OU2023109096}. For example, Tabarraei\cite{WANG2019251} developed a coupling model for dynamic fracture between finite element and peridynamic subdomains, and spurious wave reflections were effectively suppressed. Giannakeas \cite{Giannakeasone,Giannakeastwo} proposed an adaptive relocation strategy model by coupling XFEM and PD for dynamic crack branching. The coupling methods significantly reduce the computational time and cost, but the processing flow of computation will become complicated. Also, the coupling methods may cause computational errors like the ghost forces at the coupling interface which need special technology to eliminate \cite{CHEN2022115669,JIANG2020107316}.

	In the actual engineering field, there are still significant computational challenges to deal with large-scale dynamic computational problem when it involves multiple scales in time and space. In spatial problems,	an effective method to solve large-scale problems is the domain decomposition (DD) \cite{AndreaDD2005, Calvo2015DomainDM}, which splits the whole computational domain into subdomains in which the subproblems are solved independently and in parallel. The solutions are then coupled together to obtain the total domain response. DD methods are often used in coupling systems \cite{FELIPPA20013247, AKSOYLU20116498, XU2021114148, Capodaglio20221738, KLAR2023434,ZENG2022114786} including the coupling method of PD mentioned above. Several methods based on DD procedures have been proposed in recent years for the parallel solution of both static and dynamic problems. For example, Farhat and Roux \cite{FETIfar} proposed finite element tearing and interconnection (FETI), which has been widely used in the engineering. However, on their own, DD methods do not address the computational burden in time integration for transient problems. So in temporal problems, we still need a way to improve the computing efficiency. Multi-time-step (MTS) method is appropriate to achieve this goal, which allows us to adopt different time steps in different subdomains while retaining the accuracy of the original undecomposed problem overall. Belytschko et al.\cite{bely1620121008, IEwkl, IEwkl1, BELYTSCHKO1979259} proposed one of the earliest applications for considering different time integration schemes in different domains, i.e., the mixed method. Then, based on a nodal partitioning of the mixed method, an explicit multi-time-step or subcycling procedure was presented by Belytschko and Neal \cite{Bely1620260205}. Since then, more studies on MTS have been proposed \cite{BENES2015571,ZHU201836}. Daniel \cite{DANIEL19981} studied the stability of subcycling algorithms in structural dynamics. Smolinski et al.\cite{SMOLINSKI2000171, Wu2000AMS} presented the element-free Galerkin method (EFGM) for diffusion problems and an MTS integration algorithm based on the modified trapezoidal rule. 
	
	Based on the significant benefits of developments in the DD and MTS methods, more researchers began focusing on coupling MTS and DD to address problems. For example, dynamic problem can be modeled using finer spatial and temporal discretization through DD while using MTS on a time scale to reduce the computational cost \cite{BRUN201241,BRUN201219}. Prakash \cite{Prakash2004AFM,PRAKASH201451} presented an efficient and accurate MTS coupling method using FETI domain decomposition and extended it to highly nonlinear structural dynamics. Lindsay \cite{LINDSAY2016382} proposed a method for decomposing the domain into overlapping subdomains and using different time steps in different subdomains, in which the subdomain was modeled using PD. Kruis et al.\cite{BENES2018247} proposed an explicit–implicit MTS method based on FETI for parabolic problems. Bertrand and Grange \cite{GRANGE2021103604} presented a primal coupling algorithm based on a velocity gluing at the interface between two subdomains to enable consideration of the heterogeneity of both.

	In view of the development of the DD and MTS methods when facing large-scale dynamic problem, we divide the whole domain into two subdomains in this study, applying the CCM model and PD model through the Arlequin framework \cite{Arlequintool}, while adopting different time integral schemes for the different domains. Concurrently, the PD theory gives full play to its advantages in simulating and determining the damage in structurally dangerous domains. As such, the computational cost is reduced to a minimum, enabling the practical use of PD in engineering applications. We chose the classical finite element method (FEM) for the numerical computation of the CCM domain and the peridynamics-based finite element method (PeriFEM) for the PD domain \cite{RN48}, which consistently implements the framework of the classical finite element method so that it can be easily integrated into the finite element solution process of existing commercial software \cite{cmes.2023.026922}. 

	The rest of this paper is organized as follows: Section 2 introduces the proposed multi-time-step coupling of peridynamics and classical continuum mechanics model (MTS-PDCCM) in detail, including reviews of the basic formulations and discretization of PD and CCM, the coupling method in the Arlequin framework, and the MTS method. In Section 3, the proposed model is evaluated. Numerical examples are presented in Section 4. Finally, conclusions are drawn in Section 5.

	\section{Multi-time-step coupling of the PD and CCM models}
		\subsection{Basic formulations of PD and CCM}

\begin{figure}[t]
				\begin{center}
					\includegraphics[scale=0.8]{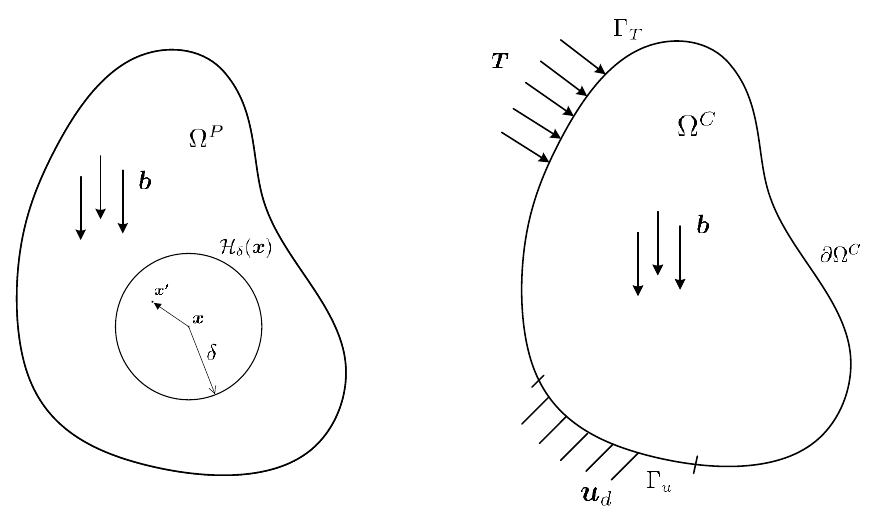}
					\caption[]{(a)the peridynamic problem, (b)the classical continuum problem.}
					\label{fig.1}
				\end{center}
			\end{figure}

			Consider an elastic body occupying $\Omega^P\subset R^d$, where $d$ is the number of space dimensions. For simplicity, we adopt the bond-based peridynamic model \cite{SILLING2000175}, which assumes	that a point $\bm{x}$ in the $\Omega^{P}$ interacts with all points in its neighborhood, $\mathcal{H}_{\delta}(\bm{x})=\left\{\bm{x}'\in \Omega^P:\left|\bm{x}'-\bm{x}\right| \leq \delta\right\} $, where $\delta$ is the peridynamic horizon denoting the cut-off radius of the action scope of $\bm{x}$, as shown in Fig.\ref{fig.1}(a). From \cite{SILLING2000175}, the peridynamic equation of motion at a point $\bm{x}$ is:
			\begin{equation}
				\label{eq.1}
				\rho(\bm{x})\ddot{\bm{u}}(\bm{x},t)=\int_{\mathcal{H}_{\delta}}\bm{f}(\bm{x}'-\bm{x})dV_{x'}+\bm{b}(\bm{x},t),
			\end{equation}
			where $\rho$ is the mass density, $\bm{u}(\bm{x},t)$ denotes the displacement field at time $t$, the dot represents the time derivative, and $\bm{b}$ represents the external body force density. The $\bm{f}$ is a pairwise force function describing the interaction between material points $\bm{x}$ and $\bm{x}'$, and $V_{x'}$ represent the volume of $\bm{x}'$. Boundary conditions cannot be directly applied in the peridynamic model due to its nonlocal property. So, constraints should be added to the model in other ways. For example, the surface tractions are considered as part of body force $\bm{b}$ in \cite{082014madenci-2}.
			
			Under the assumption of linear elasticity and small deformations, the vector-valued function $\bm{f}(\bm{\xi})$ takes the following format \cite{LI202356}:
			\begin{equation}
				\label{eq.2}
				\bm{f}(\bm{\xi})=\frac{c(\bm{x},\bm{\xi})+c(\bm{x^{\prime}},\bm{\xi})}{2}\bm{\xi}\otimes\bm{\xi}\cdot\bm{\eta}(\bm{\xi}),
			\end{equation}
			where $\bm{\xi}=\bm{x^{\prime}}-\bm{x} $ is the relative position vector called a bond, $\bm{\eta}(\bm{\xi})=\bm{u}(\bm{x^{\prime}})-\bm{u}(\bm{x}) $ is the relative displacement vector with the displacement field $ \bm{u} $, and $ c(\bm{x},\bm{\xi}) $ is the micromodulus function. For homogeneous materials, $ c(\bm{x},\bm{\xi})=c(|\bm{\xi}|), \forall \bm{x}\in\Omega^P $.
			
			The stretch-based criterion has been widely used to characterize fracture simulations. When the bond stretch $s$ is greater than a critical value $s_{crit}$, which can be derived through the energy release rate, the bond breaks irreversibly. The bond stretch $s$ is defined as:
			\begin{equation}
				\label{eq.3}
				s=\frac{|\bm{\xi}+\bm{\eta}|-|\bm{\xi}|}{|\bm{\xi}|}.
			\end{equation}
			A history-dependent scalar-valued function $\mu(\bm{\xi}, t) $ is used to describe the status of bonds:
			\begin{equation}
				\label{eq.4}
				\mu(\bm{\xi}, t)=\left\{\begin{array}{ll}
					1, & \text { if } s\left(t^{\prime}, \bm{\xi}\right)<s_{crit} \quad \text {for all}\ 0 \leqslant t^{\prime} \leqslant t, \\
					0, & \text {otherwise},
				\end{array}\right.
			\end{equation}
			where $ t $ and $ t^{\prime} $ denote the computational time steps, and the effective damage for each point $ \bm{x} $ is defined as: 
			\begin{equation}
				\label{eq.5}
				\phi(\bm{x},t)=1-\frac{\int_{H_{\delta(\bm{x})}} \mu(\bm{\xi}, t)d V_{\bm{\xi}}}{\int_{H_{\delta(\bm{x})}}  d V_{\bm{\xi}}},
			\end{equation}
			which can indicate damage to the structure.
			
			For a quasi-static problem, we do not consider the $\ddot{\bm{u}}(\bm{x},t)$ in Eq.(\ref{eq.1}). From \cite{RN48}, the finite element framework is used to solve PD problems. In this case, the total potential energy of $\Omega^P$ in Fig.\ref{fig.1}(a) can be rewritten as:
			\begin{equation}
				\label{eq.6}
				\Pi^{PD}(\bm{u})=\frac{1}{4}\int_{\bar\Omega} \bar{\bm{f}}(\bm{x^{\prime}}, \bm{x}) \cdot \bar{\bm{\eta}}(\bm{x^{\prime}}, \bm{x}) d \bar{V}_{\bm{x^{\prime}}\bm{x}} - 	\int_{\Omega^P}  \bm{b}(\bm{x})\cdot \bm{u(\bm{x})} d V\qquad \bm{x}\in \Omega^{P},
			\end{equation}
			where the first and second terms on the right-hand side are the deformation energy and external work, respectively.
			
			The integral operation in Eq.(\ref{eq.6}) is defined as \cite{RN48}:
			\begin{equation}
				\label{eq.7}
				\int_{\bar\Omega} \bar{\bm{g}}(\bm{x^{\prime}}, \bm{x}) d \bar{V}_{\bm{x^{\prime}}\bm{x}}=\int_{\Omega^P}\int_{\Omega^P} \bm{g}(\bm{\xi}) d V_{\bm{\xi}} d V_{\bm{x}},
			\end{equation}
			where $ \bar\Omega $ is an integral domain generated by two $ \Omega^P $s, and $\bar{\bm{g}}(\bm{x^{\prime}}, \bm{x})$ is a double-parameter function related to $\bm{g}(\bm{\xi})$ and is defined on $ \bar\Omega $.
		
			Similarly, we consider another elastic body occupying $\Omega^C\subset R^d$. This solid is subjected to body forces $\bm{b}$, surface tractions $\bm{T}$ over a portion $\Gamma_T$ of the boundary $\partial\Omega^C$, and displacement $\bm{u}_d$ on the boundary $\Gamma_u$, where the subscript represents the type of boundary that is constrained.
		
			According to basic knowledge of CCM, the total potential energy of $\Omega^C$ in Fig.\ref{fig.1}(b) reads as follows:
			\begin{equation}
				\label{eq.8}
					\Pi^{CCM}(\bm{u})=\frac{1}{2}\int_{\Omega^C}\bm{\sigma}(\bm{u}):\bm{\varepsilon}(\bm{u})dV-\int_{\Omega^C}\bm{b}\cdot\bm{u}dV-\int_{\Gamma_T}\bm{T}\cdot\bm{u}d\Gamma 	\qquad \bm{x}\in \Omega^C \ {\rm and }\ \bm{u}=\bm{u}_d \ {\rm on}\ \Gamma_u,
			\end{equation}
			where $\bm{\sigma}(\bm{u})$ and $\bm{\varepsilon}(\bm{u})$ are the Cauchy stress tensor and strain tensor, respectively, associated with the displacement field $\bm{u}$. The stress and strain tensors are assumed to be related through Hooke's law, and only isotropic materials are considered in this study. 
			\begin{equation}
				\label{eq.9}
				\sigma_{ij}=D_{ijhk}\varepsilon_{hk}\quad {\rm and}\quad
				D_{ijhk}=D_{jihk}=D_{hkij},
			\end{equation}
			where $i,j,h$ and $k$ range from 1 to 3.

		\subsection{Coupled model from the energy perspective}
			For CCM problems adopting the DD method in finite element tearing and interconnecting (FETI) \cite{Prakash2004AFM}, the configurational $\Omega^C$ is divided into $S$ subdomains $\Omega^I$, i.e., $\Omega^C=\Omega^1\cup\Omega^2\dots\cup\Omega^S$. Then, the total potential energy of $\Omega^C$ can be written as:
			\begin{equation}
				\label{eq.10}
				\Pi^{Total}(\bm{u})=\sum_{I}^{S}\Pi^I(\bm{u}^I).
			\end{equation} 
			Establishing the continuity constraints of the contact surface for each subdomain is necessary.
			\begin{equation}
				\label{eq.11}
				\bm{u}^I=\bm{u}^J \qquad \bm{u}^I,\bm{u}^J\in\Gamma,\quad\Gamma=\Omega^I\cap\Omega^J\quad I,J\in S.
			\end{equation}
			However, it is difficult to guarantee that Eq.(\ref{eq.11}) holds during numerical simulations. Thus, the equilibrium of the interface forces through the Lagrange multipliers in weak form is adopted to satisfy the continuity constraints, that is,
			\begin{equation}
				\label{eq.12}
				\tilde{J}_{\lambda}^{I,J}=\int_{\Gamma}(\bm{u}^I-\bm{u}^J)\bm{\lambda}^{I,J}d\Gamma\qquad \bm{u}^I,\bm{u}^J\in\Gamma,\ \Gamma=\Omega^I\cap\Omega^J\quad I,J\in S.
			\end{equation}
			Now the total potential energy is:
			\begin{equation}
				\label{eq.13}
				\tilde{\Pi}^{Total}(\bm{u})=\sum_{I}\Pi^I(\bm{u}^I)+\sum_{N}\tilde{J}_{\lambda}^{I,J},
			\end{equation}
			where $N$ is the sum of the contact surfaces between subdomains.
			
			According to the principle of minimum potential energy, the solution of Eq.(\ref{eq.13}) is also the solution of:
			\begin{equation}
				\label{eq.14}
				\mathop{\min}_{\bm{u}} \mathop{\max}_{\bm{\lambda}}\tilde{\Pi}^{Total}(\bm{u}).
			\end{equation}
			
			For the dynamic problem, the total Hamiltonian of the classical continuum domain is due to the kinetic energy and potential energy and is given by:
			\begin{equation}
				\label{eq.15}
				H^{CCM}(\bm{u},\dot{\bm{u}})=E^{CCM}(\dot{\bm{u}})+\Pi^{CCM}(\bm{u}).
			\end{equation}
			Similarly, for the peridynamic domain, the total Hamiltonian is: 
			\begin{equation}
				\label{eq.16}
				H^{PD}(\bm{u},\dot{\bm{u}})=E^{PD}(\dot{\bm{u}})+\Pi^{PD}(\bm{u}),
			\end{equation}
			where $E^{k}(\dot{\bm{u}})$ is the kinetic energy:
			\begin{equation}
				E^{k}(\dot{\bm{u}})=\int_{\Omega^{k}}\frac{1}{2}\rho\dot{\bm{u}}\cdot\dot{\bm{u}}dV,
			\end{equation}
			and $\rho$ denotes the mass density, $\dot{\bm{u}}$ is the velocity field; the superscript $k$ stands for domain, CCM or PD.
				
			For simplicity, we present a domain decomposition with $\Omega$ subdivided into two subdomains: $\Omega^C$ and $\Omega^P$ in Fig.\ref{fig.2}.
			
			According to Eq.(\ref{eq.10}), for the current coupling problems, due to the nonlocal property of the PD model, we introduce the Arlequin approach in \cite{ArlequinHan} to implement a partitioning of the energy over the overlapping domain, $\Omega^O$, using complementary weight functions. So, the total Hamiltonian of structure $\Omega$ in Fig.\ref{fig.2} reads: 
				\begin{equation}
				\label{eq.17}
				H^{Total}(\bm{u},\dot{\bm{u}})=\alpha(\bm{x}) H^{CCM}+(1-\alpha(\bm{x}))H^{PD} \quad \bm{x}\in \Omega,
			\end{equation}	
			where the weight function $\alpha(\bm{x})$ should satisfy:
			
			\begin{figure}[t]
				\begin{center}
					\includegraphics[scale=0.9]{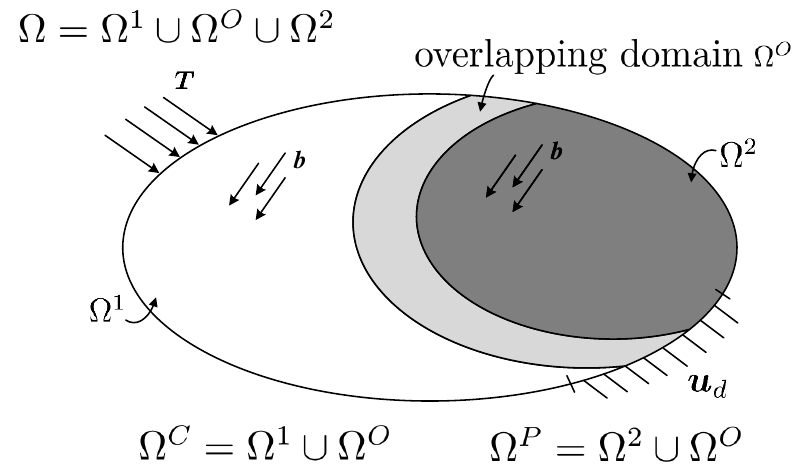}
					\caption[]{Decomposition of two subdomains.}
					\label{fig.2} 
				\end{center}
			\end{figure}

			\begin{equation}
				\label{eq.18}
				1)\ \alpha(\bm{x}) \in [0,1],\ \forall\bm{x}\in \Omega^O; \quad 2)\  \alpha(\bm{x}) =1,\ \forall\bm{x}\in \Omega^{1};\quad 3)\  \alpha(\bm{x})=0,\ \forall\bm{x}\in \Omega^{2},\ \Omega^{C}\cap\Omega^{P}=\Omega^O.
			\end{equation}
			
			However, continuity constraints are still required for the overlapping domain:
			\begin{equation}
				\label{eq.19}
				\bm{X}^{CCM}(\bm{x})=\bm{X}^{PD}(\bm{x})  \qquad \bm{x}\in\Omega^O,
			\end{equation}
			where $\bm{X}$ represents the kinematic quantities, which can be taken as displacements $\bm{u}$, velocities $\dot{\bm{u}}$, or accelerations $\ddot{\bm{u}}$.
			For the kinematic quantities, all three quantities above should be satisfied. However, from the discretization point of view, we cannot enforce the continuity of all kinematic quantities in the overlapping domain \cite{FETICG}. So, in this study, we only consider the continuity of velocities. Using the Lagrange multipliers, the weak form of continuity constraints is: 
			\begin{equation}
				\label{eq.20}
				\tilde{H}_{\lambda}^{L}=\int_{\Omega^O}(\dot{\bm{u}}^{CCM}-\dot{\bm{u}}^{PD})\bm{\lambda} dV\qquad \bm{x}\in\Omega^O.
			\end{equation}
			Thus, the total Hamiltonian of the coupled problem is: 
			\begin{equation}
				\label{eq.21}
				\tilde{H}^{Total}(\bm{u},\dot{\bm{u}})=\alpha(\bm{x}) H^{CCM}+(1-\alpha(\bm{x}))H^{PD}+\tilde{H}_{\lambda}^{L}\quad \bm{x}\in \Omega.
			\end{equation}	
			The solution of Eq.(\ref{eq.21}) is:
			\begin{equation}
				\label{eq.22}
				\mathop{\min}_{\bm{u},\dot{\bm{u}}} \mathop{\max}_{\bm{\lambda}}\tilde{H}^{Total}(\bm{u},\dot{\bm{u}}).
			\end{equation}
			To solve for the respective domains, Eq.(\ref{eq.22}) is taken apart as follows:
			\begin{subequations}
				\begin{align}
					\mathop{\min}_{\bm{u},\dot{\bm{u}}} \mathop{\max}_{\bm{\lambda}}\quad\alpha(\bm{x}) H^{CCM}+\int_{\Omega^0}\dot{\bm{u}}^{CCM}\bm{\lambda} dV\quad \bm{x}\in\Omega^{C},\\
					\mathop{\min}_{\bm{u},\dot{\bm{u}}} \mathop{\max}_{\bm{\lambda}}\quad(1-\alpha(\bm{x}))H^{PD}-\int_{\Omega^0}\dot{\bm{u}}^{PD}\bm{\lambda} dV\quad \bm{x}\in\Omega^{P}.
				\end{align}
				\label{eq.23}
			\end{subequations}
		
		\subsection{Discretization formulation}
			To facilitate computations in numerical simulations, we choose the finite element method (FEM) to discretize the CCM problem and the peridynamics-based finite element method (PeriFEM) \cite{RN48} to discretize the PD problem, of which the computational framework is consistent with FEM. So, the finite element mesh can be identical for the whole structure, eliminating the need to interpolate in the overlapping domain for different meshes, and the computing efficiency is improved.
			
			Using the interpolation technique of classical FEM, the displacement field $\bm{u}$ can be approximately expressed as:
			\begin{equation}
				\label{eq.24}
				\bm{u}(\bm{x})=\sum_{e=1}^{E^C_{all}}\bm{N}_i(\bm{x})\bm{u}_i,\qquad \bm{\varepsilon}(\bm{x})=\sum_{e=1}^{E^C_{all}}\bm{B}_i(\bm{x})\bm{u}_i,
			\end{equation}
			where $E^C_{all}$ is the total finite element number and $\bm{u}_i$ is the nodal displacement vector of element $\Omega_i$. $\bm{N}_i$ and $\bm{B}_i$ are the shape function matrix and strain matrix of element $\Omega_i$, respectively, that is,
			\begin{equation}
				\bm{N}_i(\bm{x})=
				\begin{bmatrix}
					N_{i_{1}}(\bm{x})& 0&0& N_{i_{2}}(\bm{x})& 0&0& \cdots& N_{i_{n_{i}}}(\bm{x})& 0&0\\
					0& N_{i_{1}}(\bm{x})&0& 0& N_{i_{2}}(\bm{x})&0& \cdots& 0& N_{i_{n_{i}}}(\bm{x})&0\\
					0&0& N_{i_{1}}(\bm{x})&0& 0& N_{i_{2}}(\bm{x})& \cdots& 0&0 &N_{i_{n_{i}}}(\bm{x})
				\end{bmatrix},
			\end{equation}
			\begin{equation}
				\bm{B}_i=
				\begin{bmatrix}
					\frac{\partial N_{i_{1}}}{\partial x}&  0& 0 &\cdots&\frac{\partial N_{i_{n_{i}}}}{\partial x}&  0&0\\
					0&  \frac{\partial N_{i_{1}}}{\partial y}& 0&\cdots&0&  \frac{\partial N_{i_{n_{i}}}}{\partial y}& 0\\
					0&  0& \frac{\partial N_{i_{1}}}{\partial z}&\cdots&0&  0& \frac{\partial N_{i_{n_{i}}}}{\partial z}\\
					\frac{\partial N_{i_{1}}}{\partial x}&  \frac{\partial N_{i_{1}}}{\partial y}& 0&\cdots&	\frac{\partial N_{i_{n_{i}}}}{\partial x}&  \frac{\partial N_{i_{n_{i}}}}{\partial y}& 0\\
					0&\frac{\partial N_{i_{1}}}{\partial z}&  \frac{\partial N_{i_{1}}}{\partial y}&\cdots&	0&\frac{\partial N_{i_{n_{i}}}}{\partial z}&  \frac{\partial N_{i_{n_{i}}}}{\partial y}\\
					\frac{\partial N_{i_{1}}}{\partial z}&  0& \frac{\partial N_{i_{1}}}{\partial x}&\cdots&\frac{\partial N_{i_{n_{i}}}}{\partial z}&  0& \frac{\partial N_{i_{n_{i}}}}{\partial x}
				\end{bmatrix}.
			\end{equation}
			Following the definition of $\bm{N}_i,\bm{B}_i$ and the relationship between strain and stress in Eq.(\ref{eq.9}), the element stiffness matrix of $\Omega_i$ can be written as:
			\begin{equation}
				\bm{k}_i^e=\int_{\Omega_i}\bm{B_i}^T\bm{D}\bm{B_i}dV,
			\end{equation}
			where $\bm{D}$ is the elastic matrix.
			
			In the PeriFEM, the displacement field of the peridynamic element (PE), which is generated from two finite elements \cite{RN48}, is expressed as:
			\begin{equation}
				\bm{u}(\bm{x})=\sum_{e=1}^{E^P_{all}}\bar{\bm{N}}_e(\bm{x}',\bm{x})\bar{\bm{u}}_e,
			\end{equation}
			where $E^P_{all}$ is the total PE number and $\bar{\bm{u}}_e$ is the nodal displacement vector of PE $\bar{\Omega}_e$, that is,
			\begin{equation}
				\bar{\bm{N}}_e(\bm{x'},\bm{x})=
				\begin{bmatrix}
					\bm{N}_j(\bm{x'})& \bm{0}\\
					\bm{0}& \bm{N}_i(\bm{x})
				\end{bmatrix},\quad
				\bar{\bm{u}}_e=
				\begin{bmatrix}
					\bm{u}_j(\bm{x}')\\
					\bm{u}_i(\bm{x})\\
				\end{bmatrix}.
			\end{equation}
			Also, there is a $\bar{\bm{B}}_e$ matrix called the difference matrix for the shape function given as:
			\begin{equation}
				\bar{\bm{B}}_e=\bar{\bm{H}}\bar{\bm{N}}_e(\bm{x}',\bm{x}),
			\end{equation}
			where $\bar{\bm{H}} = [\bm I\quad -\bm{I}]$ and $\bm I$ is an identity matrix of dimension $d$.
			
			In addition, the three-dimensional (3D) micromodulus tensor for homogeneous materials has the matrix form:
			\begin{equation}
				\bar{\bm{D}}(\bm{\xi})=c(|\bm{\xi}|)\mu(\bm{\xi},t)
				\begin{bmatrix}
					\xi_1^2& \xi_1\xi_2&\xi_1\xi_3\\
					\xi_2\xi_1& \xi_2^2&\xi_2\xi_3\\
					\xi_3\xi_1& \xi_3\xi_2&\xi_3^2
				\end{bmatrix}.
			\end{equation}
			At last, we have the expression of the PE stiffness matrix of $\bar{\Omega}_e$, which is:
			\begin{equation}
				\bm{k}_e^{PE}=\int_{\bar{\Omega}_e}\bar{\bm{B}}_e^T\bar{\bm{D}}\bar{\bm{B}}_ed\bar{V}\quad \bm{x},\bm{x}'\in \bar{\Omega},
			\end{equation}
			where the integral operation is defined in Eq.(\ref{eq.7}).
			
			Finally, the solution of Eq.(\ref{eq.23}) is the following linear system in time $t$:
			\begin{equation}
				\left\{
				\begin{aligned}
					\bm{M}^{FE}\ddot{\bm{u}}^{FE}_i+\bm{K^}{FE}{\bm{u}}^{FE}_i+\bm{G}^{FE^T}\bm{\lambda}_i&=\bm{P}^{FE}_i\quad \bm{x}\in\Omega^{C}\\
					\bm{M}^{PD}\ddot{\bm{u}}^{PD}_i+\bm{K}^{PD}{\bm{u}}^{PD}_i-\bm{G}^{PD^T}\bm{\lambda}_i&=\bm{P}^{PD}_i\quad \bm{x}\in\Omega^{P}\\
					\bm{G^}{FE}\dot{\bm{u}}^{FE}_i-\bm{G}^{PD}\dot{\bm{u}}^{PD}_i&=\bm{0}\quad \bm{x}\in\Omega^{O}.
				\end{aligned}
				\right.
				\label{eq.33}
			\end{equation}   
			where $\bm{M}^k$ represents the symmetric, positive-definite mass matrix, the superscript $k$ represents the subdomain, and $T$ denotes the transpose of a matrix. $\bm{G}^k$ is the Boolean connectivity matrix that extracts the corresponding quantities of the overlapping domain. $\bm{\lambda}_i$ is the vector of Lagrange multipliers force in the overlapping domain, and the subscript $i$ stands for the time step.
		
			To reduce the computation, the diagonal mass matrix is chosen as $\bm{M}^k$, and $\bm{K^}{FE},\bm{K}^{PD},\bm{P}^{FE},\bm{P}^{PD}$ can be defined as follows:
			\begin{equation}
				\label{eq.34}
				\begin{aligned}
					\bm{K}^{FE}&=\alpha (\bm{x})\sum\bar{\bm{G}}_C^T\bm{k}^e\bar{\bm{G}}_C,\\
					\bm{K}^{PD}&=(1-\alpha (\bm{x}))\sum\bar{\bm{G}}_P^T\bm{k}^{PE}\bar{\bm{G}}_P,\\
					\bm{P}^{FE}&=\alpha (\bm{x})\sum\bar{\bm{G}}_C^T\bm{N}^T\bm{b},\\
					\bm{P}^{PD}&=(1-\alpha (\bm{x}))\sum\bar{\bm{G}}_P^T\bm{N}^T\bm{b},
				\end{aligned}
			\end{equation}  
			where $\alpha (\bm{x})$ is the weight function in Eq.(\ref{eq.18}), and $\bar{\bm{G}}_C,\bar{\bm{G}}_P$ are the transform matrices of the degree of freedom for domains $\Omega^C$ and $\Omega^P$, respectively.

		\subsection{Time integration}
		When solving the discrete system in Eq.(\ref{eq.33}), a step-by-step direct time integration method for structural dynamics problems must be used to advance the solution through time. In this study, the Newmark-$\beta$ scheme \cite{newmarkb} is chosen for time integration, that is:
		\begin{equation}
			\begin{aligned}
				\dot{\bm{u}}_i&=\dot{\bm{u}}_{i-1}+\triangle t[(1-\gamma)\ddot{\bm{u}}_{i-1}+\gamma\ddot{\bm{u}}_i],\\
				\bm{u}_i&=\bm{u}_{i-1}+\triangle t\dot{\bm{u}}_{i-1}+\triangle t^2[(1/2-\beta)\ddot{\bm{u}}_{i-1}+\beta \ddot{\bm{u}}_i],
			\end{aligned}
			\label{eq.38}
		\end{equation}
		where $\bm{u}_i,\dot{\bm{u}}_i$ and $\ddot{\bm{u}}_i$ are displacements, velocities, and accelerations at time $i$, respectively. $\triangle t$ is the time step, and $\gamma$ and $\beta$ are the parameters chosen to control the stability and accuracy, respectively. This method is unstable when $\gamma<1/2$. When $\gamma\geq1/2$, this method is unconditionally stable for $\beta\geq\dfrac{1}{4}(\gamma+\dfrac{1}{2})^2$. So, two common computation choices are: i) explicit integration with $\gamma=1/2$ and $\beta =0$ (Velocity Verlet Method), and ii) implicit integration with $\gamma=1/2$ and $\beta =1/4$ (Average Acceleration Method).
		
		\subsection{The multi-time-step coupled algorithm}
\begin{figure}[t]
				\begin{center}
					\includegraphics[scale=0.8]{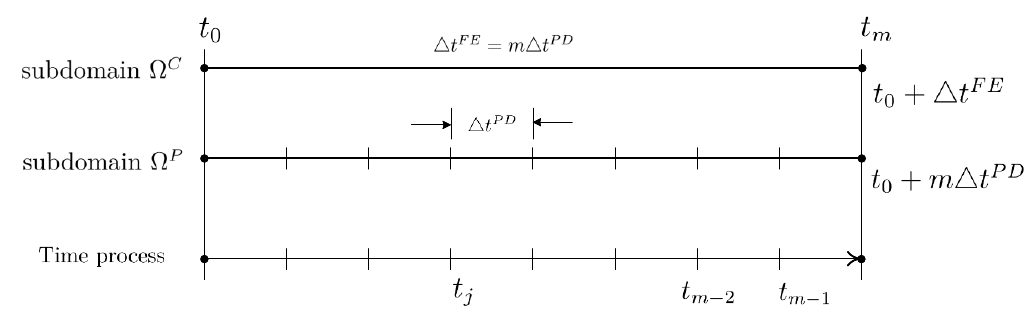}
					\caption[]{Time steps for the two subdomains with time step ratio $m$, where $t_j=t_0+j\triangle t^{PD},\  j=0,\cdots,m$.}
					\label{fig.3} 
				\end{center}
			\end{figure}

			For simplicity, we still consider decomposing $\Omega$ into two subdomains, $\Omega^C$ and $\Omega^P$. When solving Eq.(\ref{eq.33}) by time integration Eq.(\ref{eq.38}), for subdomains $\Omega^C$ and $\Omega^P$, we choose different time steps $\triangle t^{FE}$ and $\triangle t^{PD}$, respectively, where $\triangle t^{FE}=m\triangle t^{PD}$ and $m$ is an integer representing the time step ratio. For notational simplicity, the coupling method for advancing the solution to $\triangle t^{FE}$ from $t_0$ to $t_m = t_0 + \triangle t^{FE}$ is shown in Fig.\ref{fig.3}. This can be easily generalized to advance the solution from a known state $t_n$ to $t_{n+m}$. The Newmark-$\beta$ parameters for the two subdomains are ($\gamma^{FE}$,$\beta^{FE}$) and ($\gamma^{PD}$,$\beta^{PD}$), respectively.
			
			The fully discretized linear equations for subdomain $\Omega^{C}$ are:
			\begin{subequations}
				\label{eq.39}
				\begin{align}
					\bm{M}^{FE}\ddot{\bm{u}}^{FE}_m+\bm{K^}{FE}{\bm{u}}^{FE}_m+\bm{G}^{FE^T}\bm{\lambda}_m=\bm{P}^{FE}_m,\\
					\dot{\bm{u}}_m^{FE}=\dot{\bm{u}}_{0}^{FE}+\triangle t^{FE}[(1-\gamma^{FE})\ddot{\bm{u}}_{0}^{FE}+\gamma^{FE}\ddot{\bm{u}}_m^{FE}],\\
					\bm{u}_m^{FE}=\bm{u}_{0}^{FE}+\triangle t^{FE}\dot{\bm{u}}_{0}^{FE}+(\triangle t^{{FE}})^2[(1/2-\beta^{FE})\ddot{\bm{u}}_{0}^{FE}+\beta^{FE} \ddot{\bm{u}}_m^{FE}],
				\end{align}
			\end{subequations}
			where the second and third equations are obtained from Newmark-$\beta$ in Eq.(\ref{eq.38}). Meanwhile, for the subdomain $\Omega^{P}$, there should be $m$ times equations, that is, $\forall j=1,2,\cdots,m$
			\begin{subequations}
					\label{eq.40}
					\begin{align}
						\label{eq.40a}
					\bm{M}^{PD}\ddot{\bm{u}}^{PD}_j+\bm{K^}{PD}{\bm{u}}^{PD}_j-\bm{G}^{PD^T}\bm{\lambda}_j=\bm{P}^{PD}_j,\\
					\dot{\bm{u}}_j^{PD}=\dot{\bm{u}}_{j-1}^{PD}+\triangle t^{PD}[(1-\gamma^{PD})\ddot{\bm{u}}_{j-1}^{PD}+\gamma^{PD}\ddot{\bm{u}}_j^{PD}],\\
					\bm{u}_j^{PD}=\bm{u}_{j-1}^{PD}+\triangle t^{PD}\dot{\bm{u}}_{j-1}^{FE}+(\triangle t^{{PD}})^2[(1/2-\beta^{PD})\ddot{\bm{u}}_{j-1}^{PD}+\beta^{PD} \ddot{\bm{u}}_j^{PD}].
				\end{align}
			\end{subequations}
			At the final time $t_m$, the continuity of velocities in the overlapping domain can be expressed as:
			\begin{equation}
				\label{eq.41}
				\bm{G^}{FE}\dot{\bm{u}}^{FE}_m-\bm{G}^{PD}\dot{\bm{u}}^{PD}_m=\bm{0}.
			\end{equation}
			In fact, the problems can be solved directly without any additional conditions when $m=1$. However, when $m$ is an integer greater than one, the $\bm{\lambda}_j$ in Eq.(\ref{eq.40a}) is unknown at first. So, we choose a linearly interpolated quantity from the known quantities at common time steps $j=0$ and $m$, given as:
			\begin{equation}
				\label{eq.42}
				\bm{z}^{FE}_j=(1-j/m)\bm{z}_0^{FE}+(j/m)\bm{z}_m^{FE},\quad j=1,2,\cdots,m-1,
			\end{equation}
			where $\bm{z}^{FE}_j=\{\ddot{\bm{u}}^{FE}_j,\dot{\bm{u}}^{FE}_j,\bm{u}_j^{FE}\}^T$.
			
			To facilitate the derivation of the solution of Eq.(\ref{eq.39}-\ref{eq.42}), the Newmark-$\beta$ time stepping scheme is reconsidered \cite{Prakash2004AFM}, and the simpler form of the problem can be rewritten as:
			\begin{subequations}
				\begin{align}
					\label{eq.43a}
					\mathbb{M}^{FE}\mathbb{U}^{FE}_m+\mathbb{N}^{FE}\mathbb{U}^{FE}_0+\mathbb{G}^{FE}\bm{\lambda}_m&=\mathbb{P}_m^{FE},\\
					\label{eq.43b}
					\mathbb{M}^{PD}\mathbb{U}^{PD}_j+\mathbb{N}^{PD}\mathbb{U}^{PD}_{j-1}+\mathbb{G}^{PD}\bm{\lambda}_j&=\mathbb{P}_j^{PD},\\
						\label{eq.43c}
					\bm{G^}{FE}\dot{\bm{u}}^{FE}_m-\bm{G}^{PD}\dot{\bm{u}}^{PD}_m&=0,\\
					\mathbb{U}^{FE}_j=(1-j/m)\mathbb{U}_0^{FE}+(j/m)&\mathbb{U}_m^{FE},\quad j=1,2,...,m-1,
				\end{align}	
				\label{eq.43}
			\end{subequations}   
			where
			\begin{equation}
				\mathbb{M}^k=\begin{bmatrix}
					\bm{M}^k& \bm{0}&\bm{K}^k\\
					-\gamma^k\triangle t^k\bm{I}& \bm{I}&\bm{0}\\
					-\beta^k(\triangle {t^k})^2\bm{I}& \bm{0}&\bm{I}\\
				\end{bmatrix},\quad
				\mathbb{N}^k=\begin{bmatrix}
					\bm{0}& \bm{0}&\bm{0}\\
					-(1-\gamma^k)\triangle t^k\bm{I}& -\bm{I}&\bm{0}\\
					-(1/2-\beta^k)(\triangle {t^k})^2\bm{I}& -\triangle t^k\bm{I}&-\bm{I}\\
				\end{bmatrix},
			\end{equation}
			\begin{equation}
				\mathbb{G}^k=\begin{bmatrix}\bm{G}^{k^T}\\\bm{0}\\\bm{0}\\\end{bmatrix},\quad
				\mathbb{U}^k_j=\begin{bmatrix}\ddot{\bm{u}}^k_j\\\dot{\bm{u}}^k_j\\\bm{u}^k_j\\\end{bmatrix},\quad
				\mathbb{P}^k_j=\begin{bmatrix}\bm{P}^k_j\\	\bm{0}\\\bm{0}\\	\end{bmatrix},
			\end{equation}
			where the superscript $k$ denotes the subdomain and the subscript $j$ denotes the time steps.
			
			Now, for the MTS coupling problem, we can solve Eq.(\ref{eq.43}) to access the result of discrete equation on a time scale. However, in the numerical computation, we will not directly solve Eq.(\ref{eq.43}) because each equation contains more than two related unknowns, which will require significant computing resources. The equations can be solved following an alternative direct solution of decoupling each subdomain and solving them concurrently \cite{Prakash2004AFM}.
			
			The main idea is to split the kinematic quantities into two parts:
			\begin{equation}
				\mathbb{U}_m=\mathbb{V}_m+\mathbb{W}_m,
			\end{equation}
			Then, the Eq.(\ref{eq.43a}) can be split as:
			\begin{subequations}
				\begin{align}
					\label{eq.47a}
					\mathbb{M}^{FE}\mathbb{V}^{FE}_m+\mathbb{N}^{FE}\mathbb{U}^{FE}_0&=\mathbb{P}_m^{FE},\\
					\mathbb{M}^{FE}\mathbb{W}^{FE}_m+\mathbb{G}^{FE}\bm{\lambda}_m&=\bm{0},
				\end{align}
			\end{subequations}
			where $\mathbb{V}^{FE}_m$ is only computed from the external forces, which can be seen as free problems, and $\mathbb{W}^{FE}_m$ is only computed from the Lagrange multipliers force, which can be seen as link problems.
			
			Similarly, for Eq.(\ref{eq.43b}):
			\begin{subequations}
				\begin{align}
					\label{eq.48a}
					\mathbb{M}^{PD}\mathbb{V}^{PD}_j+\mathbb{N}^{PD}\mathbb{V}^{PD}_{j-1}&=\mathbb{P}_j^{PD} - \mathbb{G}^{PD}\bm{S}_j,\\
					\label{eq.48b}
					\mathbb{M}^{PD}\mathbb{W}^{PD}_j+\mathbb{N}^{PD}\mathbb{W}^{PD}_{j-1}+\mathbb{G}^{PD}(\frac{j}{m})\bm{\lambda}_m&=\bm{0} \quad  \forall j \in [1,2,...m],
				\end{align}
			\end{subequations}
			where $\mathbb{V}_0^{PD}=\mathbb{U}_0^{PD}, \mathbb{W}_0^{PD}=\bm{0}, \bm{S}_j=(1-j/m)\bm{\lambda}_0+\bm{G}^{FE}[\bm{P}_j^{FE}-(1-j/m)\bm{P}_0^{FE}-(j/m)\bm{P}^{FE}_m]$, and $\bm{\lambda}_0$ are the Lagrange multipliers force in the initial state.
			
			$\mathbb{V}_m^{FE}$ and $\mathbb{V}_m^{PD}$ can be directly solved from Eq.(\ref{eq.47a})(\ref{eq.48a}). Since $\bm{\lambda}_m$ is initially unknown, we can first solve:
			\begin{equation}
				\label{eq.unit}
				\begin{aligned}
				\mathbb{M}^{FE}\mathbb{Y}^{FE}_m+\mathbb{G}^{FE}&=\bm{0},\\	\mathbb{M}^{PD}\mathbb{Y}^{PD}_j+\mathbb{N}^{PD}\mathbb{Y}^{PD}_{j-1}+(\frac{j}{m})\mathbb{G}^{PD}&=\bm{0} \quad  \forall j \in [1,2,...m],
				\end{aligned}
			\end{equation}
			 which gives the structural response under unit load $\mathbb{Y}_m^{FE}$ and $\mathbb{Y}_m^{PD}$. Then, according to the velocity continuity condition, i.e., Eq.(\ref{eq.43c}),
			\begin{equation}
				\label{eq.49}
				[\mathbb{G}^{FE}\dot{\bm{Y}}^{FE}_m+\mathbb{G}^{PD}\dot{\bm{Y}}^{PD}_m]\bm{\lambda}_m=[\mathbb{G}^{FE}\dot{\bm{V}}^{FE}_m+\mathbb{G}^{PD}\dot{\bm{V}}^{PD}_m],
			\end{equation}
			which gives the $\bm{\lambda}_m$ and the final result of the systems as: 
			\begin{equation}
				\mathbb{U}^{FE}=\mathbb{V}^{FE}_m+\mathbb{Y}^{FE}_m\bm{\lambda}_m,\qquad\mathbb{U}^{PD}=\mathbb{V}^{PD}_m+\mathbb{Y}^{PD}_m\bm{\lambda}_m.
			\end{equation}
			This is equal to solving the system Eq.(\ref{eq.43}) but is more efficient and requires less computation.
			
			Notes about broken bonds: whether the final bond length exceeds the critical elongation results from the combination of the external load and the interface reaction of the structure. However, the interface reaction is finally solved using Eq.(\ref{eq.49}). Thus, in this study, we only consider the effect of Eq.(\ref{eq.48a}) to update the bond state and regard Eq.(\ref{eq.48b}) as elastic, that is, no broken bond.
			
			The pseudocode for the MTS coupling of the PD and CCM methods is shown in Algorithm 1.
			\begin{algorithm}[H]
				\caption{MTS coupling of PD and CCM models}
				\begin{algorithmic}[1]
					\STATE{Preparatory stage (Eq.(\ref{eq.34}) for $\mathbb{M}^{FE}$ and $\mathbb{M}^{PD}$)}\\
					\STATE{$\rhd$ Solve the Eq.(\ref{eq.unit}) under unit load for $\mathbb{Y}_m^{FE},\mathbb{Y}_m^{PD}$}\\
					\STATE{\quad Solve $\mathbb{M}^{FE}\mathbb{Y}^{FE}_m+\mathbb{G}^{FE}=\bm{0}$ }\\
					\STATE{\quad\textbf{for} $j=1:m$}\\
					\STATE{\qquad Solve $	\mathbb{M}^{PD}\mathbb{Y}^{PD}_j+\mathbb{N}^{PD}\mathbb{Y}^{PD}_{j-1}+(\frac{j}{m})\mathbb{G}^{PD}=\bm{0} $}\\
					\STATE{$\rhd$ In the following, subscript 0 refers to time $t-\triangle t^{FE}$, and $m$ refers to time $t$.}\\
					\STATE{\quad\textbf{for} $t=t_0:\triangle t^{FE}:t_{final}$}\\
					\STATE{\qquad Solve $\mathbb{M}^{FE}\mathbb{V}^{FE}_m+\mathbb{N}^{FE}\mathbb{U}^{FE}_0=\mathbb{P}_m^{FE}$}\\
					\STATE{\qquad \textbf{for} $j=1:m$}\\
					\STATE{\qquad\quad Solve $\mathbb{M}^{PD}\mathbb{V}^{PD}_j+\mathbb{N}^{PD}\mathbb{V}^{PD}_{j-1}=\mathbb{P}_j^{PD} - \mathbb{G}^{PD}\bm{S}_j$}\\
					\STATE{\qquad\quad \textbf{if} bond broken \textbf{then} }\\
					\STATE{\qquad\qquad Update bond state and $\mathbb{M}^{PD}$}\\
					\STATE{\qquad$\rhd$ Solve the Lagrange multipliers force $\bm{\lambda}_m$ at time $t_m$.}\\
					\STATE{\qquad\quad Solve $[\mathbb{G}^{FE}\dot{\bm{Y}}^{FE}_m+\mathbb{G}^{PD}\dot{\bm{Y}}^{PD}_m]\bm{\lambda}_m=[\mathbb{G}^{FE}\dot{\bm{V}}^{FE}_m+\mathbb{G}^{PD}\dot{\bm{V}}^{PD}_m]$}\\
					\STATE{\qquad $\rhd$ Update the final result in time $t_m$}\\
					\STATE{\qquad\quad $\mathbb{U}^{FE}=\mathbb{V}^{FE}_m+\mathbb{Y}^{FE}_m\bm{\lambda}_m$}
					\STATE{\qquad\quad$\mathbb{U}^{PD}=\mathbb{V}^{PD}_m+\mathbb{Y}^{PD}_m\bm{\lambda}_m$}
					\STATE{\qquad\textbf{if} bond broken \textbf{then} }\\
					\STATE{\qquad$\rhd$ Update $\mathbb{Y}_m^{PD}$ for next time iteration using steps 4 and 5}\\
				\end{algorithmic}
			\end{algorithm}

	\section{Evaluation of the coupled model}
		\subsection{Stability analysis}
		
		In this section, we will briefly demonstrate the stability of the proposed MTS coupling method through the energy partition. The change of energy for subdomains CCM and PD during a large time step $\triangle t^{FE}$ is: 
		\begin{equation}
			\mathscr{E}=\mathscr{E}_0^{FE}+\sum_{j=1}^{m}\mathscr{E}^{PD}_{j-1},
		\end{equation}
		where
		\begin{equation}
			\label{eq.52}
			\mathscr{E}_0^{FE}=-(\gamma^{FE}-1/2)(\ddot{\bm{u}}_m^{FE}-\ddot{\bm{u}}_0^{FE})^T\bm{A}^{FE}(\ddot{\bm{u}}_m^{FE}-\ddot{\bm{u}}_0^{FE})+E_\lambda^{FE},
		\end{equation}
		\begin{equation}
				\label{eq.53}
			\mathscr{E}_{j-1}^{PD}=-(\gamma^{PD}-1/2)(\ddot{\bm{u}}_{j}^{PD}-\ddot{\bm{u}}_{j-1}^{PD})^T\bm{A}^{PD}(\ddot{\bm{u}}_j^{PD}-\ddot{\bm{u}}_{j-1}^{PD})+E_\lambda^{PD},
		\end{equation}
		and 
		\begin{equation}
			\bm{A}^{k}=\bm{M}^{k}+(\triangle t^{k})^2(\beta^{k}-\gamma^{k}/2)\bm{K}^{k},
		\end{equation}
		the superscript $k$ denotes the subdomains CCM or PD.
	
		For the numerical method adopted in this study, the mass and stiffness matrices $\bm{M}^k$ and $\bm{K}^k$ are positive-definite. So, the stability of the numerical algorithm is controlled by the second term in Eqs.(\ref{eq.52}) and (\ref{eq.53}), that is, 
		\begin{equation}
			\begin{aligned}
				E_\lambda=E_\lambda^{FE}&+\sum_{j=1}^{m}E_\lambda^{PD}\\
				=1/\triangle t^{FE}(\dot{\bm{u}}^{FE}_m-\dot{\bm{u}}^{FE}_0)^T\bm{G}^{FE}(\bm{\lambda}_m^{FE}-\bm{\lambda}_0^{FE})
				&+1/\triangle t^{PD}\sum_{j=1}^{m}(\dot{\bm{u}}^{PD}_j-\dot{\bm{u}}^{PD}_{j-1})^T\bm{G}^{PD}(\bm{\lambda}_j^{PD}-\bm{\lambda}_{j-1}^{PD}).
			\end{aligned}
		\end{equation}
		The velocity during $\triangle t^{FE}$ is already constrained in Eq.(\ref{eq.43c}). Therefore, the stability of this method is controlled by the Newmark-$\beta$ time integration; that is, as long as the Newmark-$\beta$ method guarantees the stability of numerical integration, the coupling method is also stable. More details of the stability proof are given in \cite{Prakash2004AFM}. 
	
		\subsection{Error analysis}
		In the previous coupling method \cite{LINDSAY2016382}, the author analyzed the influence of truncation error for results analytically and numerically for the MTS coupling of PD models. In this study, we adopt the same MTS coupling scheme, so the truncation error is similar. Since the coupling method is based on the Arlequin framework, the error analysis in this study focuses on the influence of overlapping domain width and the types of weight functions. By comparison with the results from FEM, the optimal processing method for the overlapping domain is selected to achieve efficient computation considering the accuracy.
		
\begin{figure}[h]
			\begin{center}
				\includegraphics[scale=1.0]{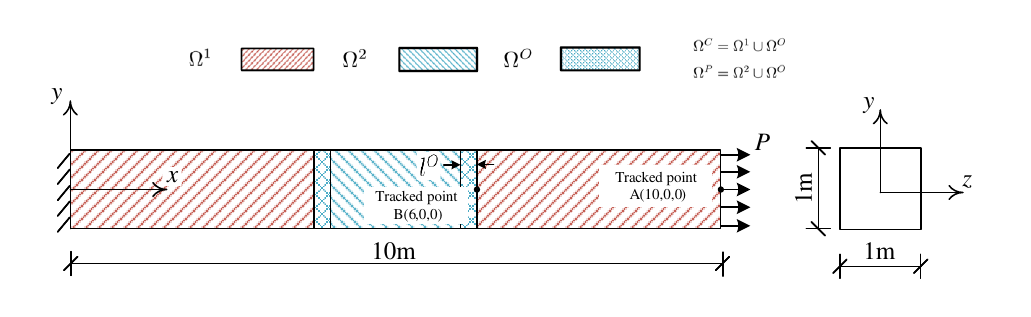}
				\caption[]{The geometry and domain decomposition, where the overlapping domain $\Omega^O=\Omega^P\cap\Omega^C$.}
				\label{fig.4} 
			\end{center}
		\end{figure}

		In this section, we choose a 3D beam loaded with tractions at the end. The geometry of the beam is $1\times1\times10$ m, the left end is fixed, and the right end is the free surface loaded at uniform traction $P=4$ MPa, as shown in Fig.\ref{fig.4}, with elastic modulus $E=6.5$ GPa, Poisson's ratio $\nu=1/4$, and density $\rho=2235$ kg/m$^3$. The beam is discretized with a uniform size $\triangle x=0.05$ m using a hexahedral element, and two tracked points are selected; one is in the center of the right free surface, A(10,0,0), and the other is located in the overlapping domain, B(6,0,0).
		
		For PD in numerical computations, the horizon $\delta =3.03\triangle x$. For three dimensions, the micromodule function in Eq.(\ref{eq.2}) is set as exponential the form \cite{cmes.2023.026922}:
		\begin{equation}
			c(|\bm{\xi}|)=c^0e^{-\frac{|\bm{\xi}|}{l}},\quad c^0=\frac{3E}{\pi \int_{0}^{\delta}r^6e^{-\frac{r}{l}}dr},
		\end{equation}
		where $l$ is the length scale parameter fixed as $l=\delta/16$ in this benchmark example.
		
\begin{figure}[t]
			\begin{center}
				\includegraphics[scale=0.8]{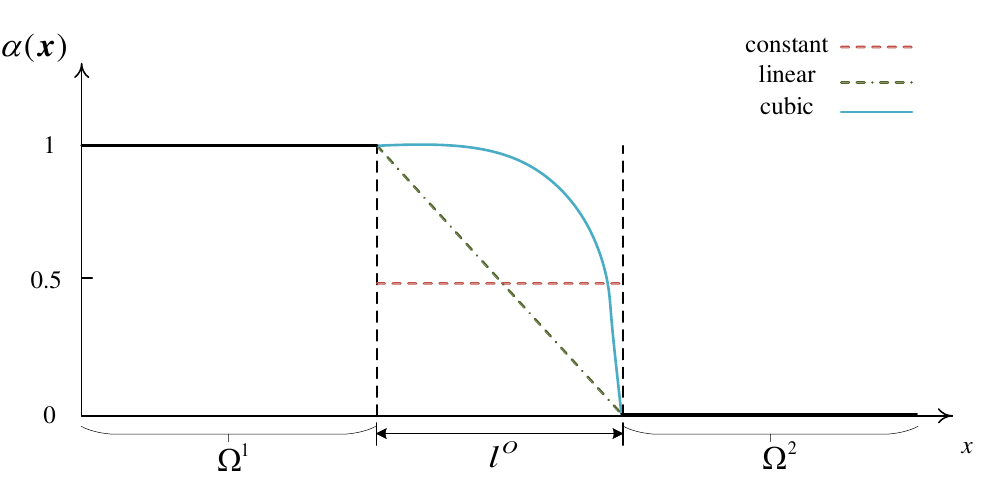}
				\caption[]{The variation of the weight function $\alpha(\bm{x})$ between subdomains.}
				\label{fig.5} 
			\end{center}
		\end{figure}

		The domain decomposition is shown in Fig.\ref{fig.4}, where the subdomain of PD is fixed as $\{\Omega^P|4<x<6\}$, and the range of CCM is $\Omega^C=\Omega\backslash\Omega^2$, where $\{\Omega|0<x<10\}$ and $\{\Omega^2|4+l^O<x<6-l^O\}$, and $l^O$ is the width of the overlapping domain.
		In selecting the weight function $\alpha(\bm{x})$ in Eq.(\ref{eq.18}), we mainly consider the constant, linear, and cubic relations as follows:
		\begin{equation}
			\alpha(\bm{x})=
			\left\{
			\begin{aligned}
				&\dfrac{1}{2}&\quad &{\rm constant}\\
				&\dfrac{l_1}{l^O}&\quad& {\rm linear},\quad \bm{x}\in \Omega^O,\\
				&(\dfrac{l_1}{l^O})^3&\quad &{\rm cubic}
			\end{aligned}
			\right.
		\end{equation}
		where $\Omega^O$ is the overlapping domain, and $l_1$ is the distance of $\bm{x}$ to the $\Omega^{FE}$ boundary. Fig.\ref{fig.5} shows the variation of the weight function between subdomains.
	
		The results of errors in displacement for tracked point A and computation parameters are provided in Table 1. The integration parameter $\gamma$ of Newmark-$\beta$ is 1/2 for all cases.
	
		First, cases 1 and 2 correspond to the undecomposed FE simulations with implicit and explicit integrations in time step $\triangle t =2.5\times10^{-5}$ s. Through this time step, the implicit and explicit results are the same, which is considered convergent, so the results of cases 1 and 2 are regarded as the reference for the following comparison. The global error is calculated using the following formula:
		\begin{equation}
			\bm{u}_x^{err}=\frac{\sqrt{\sum_1^N||u_i^{ref}-u_i||^2}}{\sqrt{\sum_1^N||u_i^{ref}||^2}},
		\end{equation}
		where $\bm{u}$ represents the displacement of a point on the structure, $i$ is the time step, and $N$ is the total time step.
		
		\begin{table}[th]
			\renewcommand\arraystretch{1.0}
			\begin{center}
				\caption{Computation parameters and errors for different cases }
				\begin{tabular}{c|c|c|c|c|c|c |c | c }
					\hline
					\multirow{2}{*}{method}&\multirow{2}{*}{case}&\multirow{2}{*}{description}&\multirow{2}{*}{$l^O$}&\multirow{2}{*}{$\beta^{FE}$}&\multirow{2}{*}{$\beta^{PD}$}& \multirow{2}{*}{$\dfrac{\triangle t^{PD}({\rm s})}{2.5\times10^{-5}}$}& \multirow{2}{*}{m} &  \multirow{2}{*}{\makecell[c]{$\bm{{u}}_x^{err}$(\%) \\of point A}}\\				
					& & & & & & &  &\\
					\hline
					\multirow{2}{*}{\makecell[c]{undecomposed \\FEM (ref.)}} &1 &implicit  integration& \multirow{2}{*}{-} &	\multicolumn{2}{c|}{ $\beta=1/4$}  &\multicolumn{2}{c|}{\multirow{2}{*}{$\triangle t=2.5\times10^{-5}$}} &0.00 \\
					& 2 &explicit integration &  &  \multicolumn{2}{c|}{ $\beta=0$ } & \multicolumn{2}{c|}{}&    0.00\\
					\hline
					\multirow{10}{*}{\makecell[c]{decomposed\\PDCCM}} &3 &\multirow{3}{*}{\makecell[c]{$\alpha(\bm{x})$ type: constant}}  & $\triangle x$& \multirow{3}{*}{ $1/4$} & \multirow{3}{*}{ $1/4$}&\multirow{3}{*}{1} & \multirow{3}{*}{1}& 1.74\\
					& 4& &$3\triangle x$ &  & &&&1.72  \\
					& 5 && $5\triangle x$&  & &&&1.68 \\
					\cline{2-9}
				    &6 &\multirow{3}{*}{\makecell[c]{$\alpha(\bm{x})$ type: linear}}  & $\triangle x$& \multirow{3}{*}{ $1/4$} & \multirow{3}{*}{ $1/4$}&\multirow{3}{*}{1} & \multirow{3}{*}{1}&1.74 \\
					& 7& &$3\triangle x$ &  & &&&1.09  \\
					& 8 && $5\triangle x$&  & &&&0.92 \\
					\cline{2-9}
				    &9 &\multirow{3}{*}{\makecell[c]{$\alpha(\bm{x})$ type: cubic}}  & $\triangle x$& \multirow{3}{*}{ $1/4$} & \multirow{3}{*}{ $1/4$}&\multirow{3}{*}{1} & \multirow{3}{*}{1}&1.51\\
					& 10& &$3\triangle x$ &  & &&&0.86  \\
					& 11 && $5\triangle x$&  & &&&0.73\\
					\cline{2-9}
					& 12 &\makecell[c]{$\alpha(\bm{x})$ type: cubic\\explicit integration}& $3\triangle x$& { $1/4$} & 0& 1& 1 &0.89 \\
					\hline
					\multirow{6}{*}{MTS-PDCCM} &13 &\multirow{6}{*}{\makecell[c]{$\alpha(\bm{x})$ type: cubic\\Elastic dynamics}}  & \multirow{6}{*}{$3\triangle x$}& \multirow{3}{*}{ $1/4$} & $0$&1/2 & 2& 0.96   \\
					&14& & &  &  $1/6$&1/2 &2 &0.95  \\
					&15& & &  &  $1/4$& 1/2&2 &  0.95 \\
					\cline{5-9}
					&16& && 0&  0& 1/2&2 & 1.05\\
					\cline{5-9}
					&17& & & 1/4& 0& 1/5 & 5& 0.96\\
					&18& & &1/4 & 0& 1/10& 10&0.97\\
					\hline
				\end{tabular}
			\end{center}
		\end{table}
		
		Since the coupling method in this study is based on the Arlequin framework, the type of weight function and width of the overlapping domain are our focus. First, we consider the results of the decomposition problem, i.e., the time ratio $m=1$. Three types of weight functions are selected for comparison. In cases 3, 4, and 5, we find that in the case of the constant weight function, even if we increase the width of the overlapping domain, there is still an error of about $1.7\%$. The error is the reduction of stiffness at the boundary of the coupling subdomain owing to the absence of elements in the horizon caused by the nonlocality of the PD model. Presently, there are many studies and methods to modify the boundary effect of PD, such as the fictitious node method and the variable horizon method \cite{LIU202278,REN2017762,Oterkus10.11}. However, this study focuses on the MTS method, and to simplify the research, these processing methods are not introduced.
		Besides, we also choose the linear and cubic weight functions to reduce the stiffness weakening of the PD model in the coupling domain. By comparing the results of cases 6 to 11, it can be seen that cubic weight functions cause less error than linear weight functions. The width of the overlapping domain is also important when considering computational efficiency. We found that when the width of the overlapping domain is $3\triangle x$ in cubic cases, the error compared with that in the FEM can be reduced to less than $1\%$. Finally, to verify the results in the case of explicit integration, we calculate case 12 to prove that both explicit and implicit integration can meet the accuracy requirements. Therefore, in the following MTS example, all the weight functions are selected as the cubic form, and the width of the overlapping domain is $l^O=3\triangle x$, considering the computational efficiency.

		\begin{figure}[H]
			\begin{center}
				\includegraphics[scale=0.4]{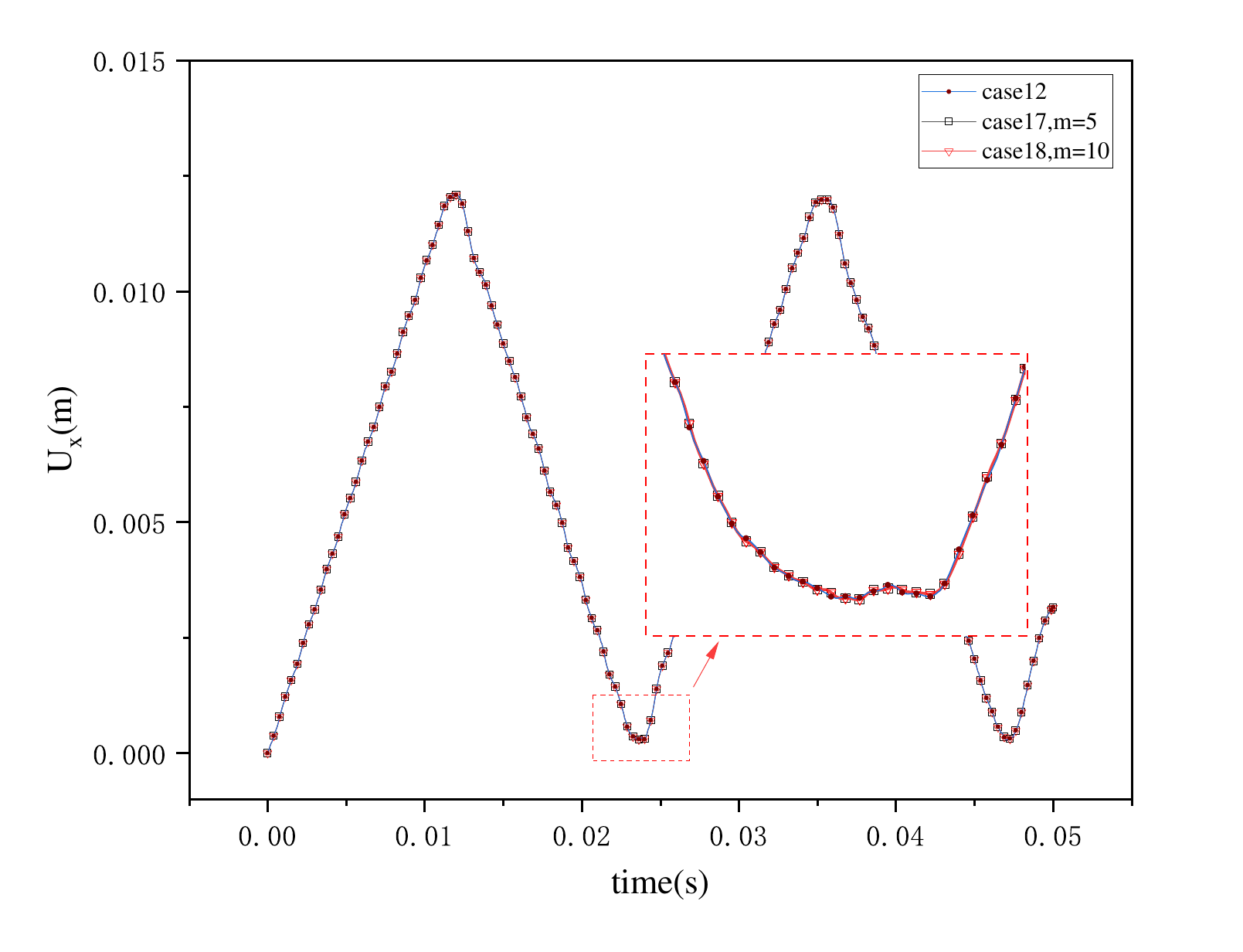}
				\caption[]{Displacement of point A in the x-direction.}
				\label{fig.6} 
			\end{center}
		\end{figure}
		In cases 13 to 15 of MTS-PDCCM, we study the influence of the integration parameter $\beta$ on the results in the PD subdomain by fixing other parameters. Notably, the result is not affected regardless of whether explicit integration or implicit integration is adopted. However, while explicit integration does not need to solve linear equations, implicit integration does. Therefore, the computational costs, which will be examined in the next section, are significantly different. Then, in case 16, we change the integration methods of the two subdomains to explicit ones. Based on the error comparison in case 12, it is still considered to be under control. Finally, we simulate the case of implicit integration in the CCM subdomain and explicit integration in the PD subdomain with time ratios of $m=5$ and $m=10$ in cases 17 and 18. Fig.\ref{fig.6} shows the x-direction displacement diagram of tracked point A in cases 12, 17, and 18, and the results are observed to be the same.
			
\begin{figure}[H]
			\centering	
			\subfigure[case17: $\triangle t=0.5\times 10^{-5}$ s]{
		\includegraphics[scale=0.35]{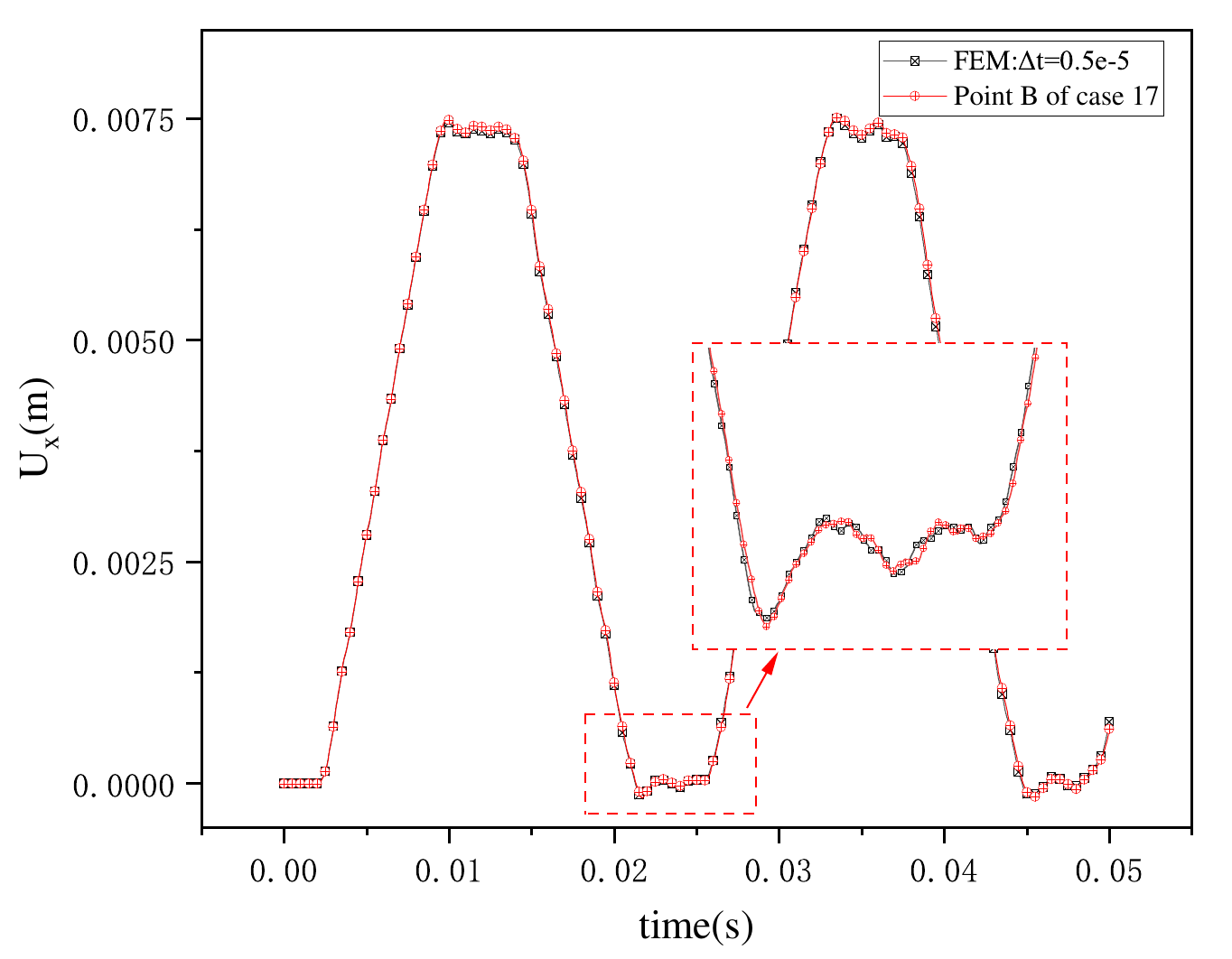}	
		}
		\quad
				\subfigure[case18: $\triangle t=2.5\times 10^{-6}$ s]{
		\includegraphics[scale=0.35]{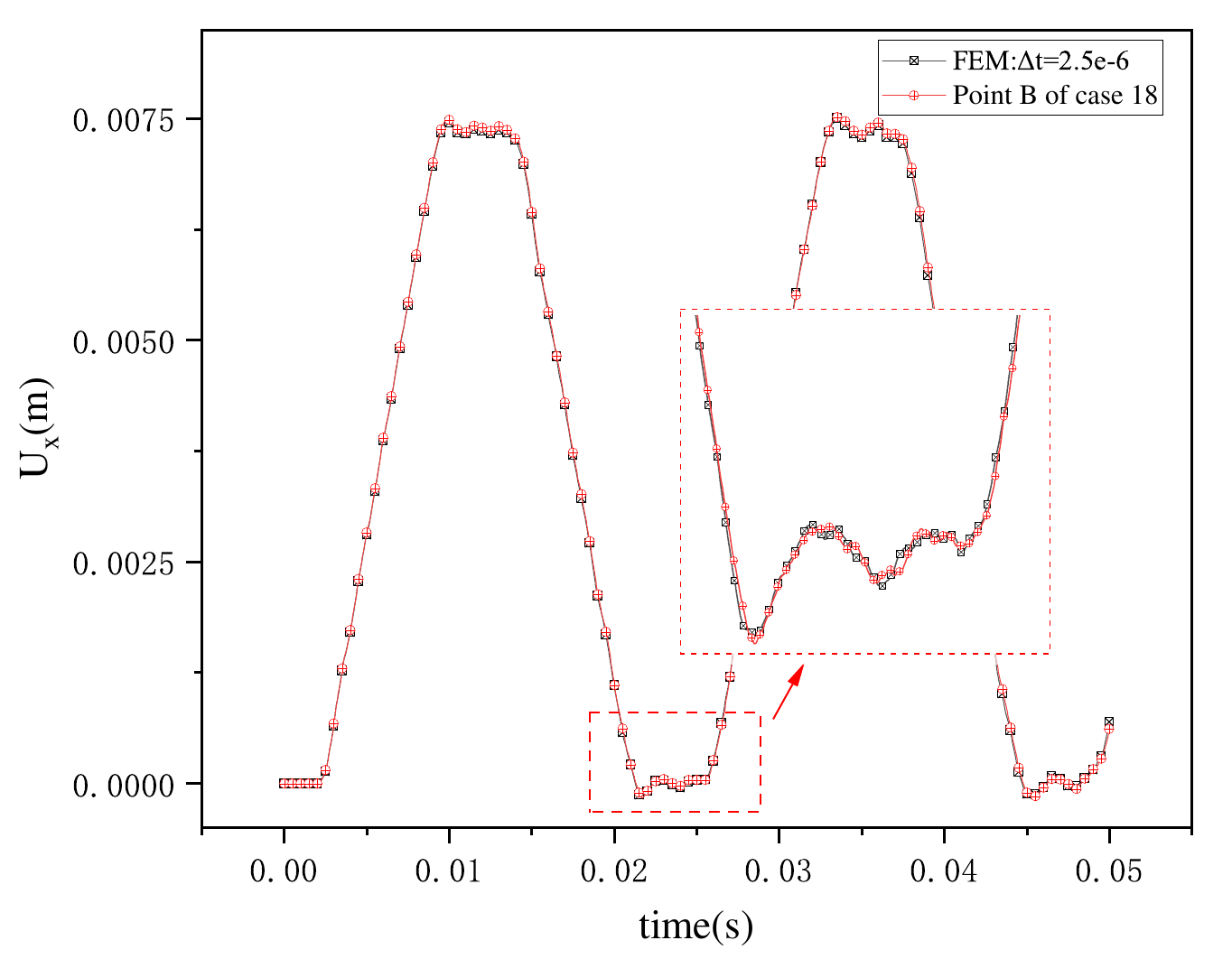}	
		}
		\caption{Displacement of point B in the x-direction}
			\label{fig.7} 
		\end{figure}

		Then, for tracked point B in the overlapping domain, we use explicit integration to compute the results of the undecomposed FEM with time steps of $\triangle t=0.5\times 10^{-5}$ s and $\triangle t=0.25\times 10^{-5}$ s to compare the results in cases 17 and 18, respectively. As shown in Fig.\ref{fig.7}, the displacement results are the same with the proposed coupling method.
		
		\subsection{Efficiency analysis}
		In this section, we analyze and compare the efficiency of simulating structural damage using the proposed MTS-PDCCM coupled and pure PD models. First of all, in the dynamic PD damage simulation, once the pre-stage is ready, the main computational requirement is solving the unknown state $t_{n+1}$ according to the discretized motion equation and updating the bond state and the total stiffness matrix. By obtaining the calculation result of $t_{n+1}$ and taking it as the initial state of the next time step, we can move forward in time until the computation is completed.
		
\begin{figure}[h]
			\begin{center}
				\includegraphics[scale=0.9]{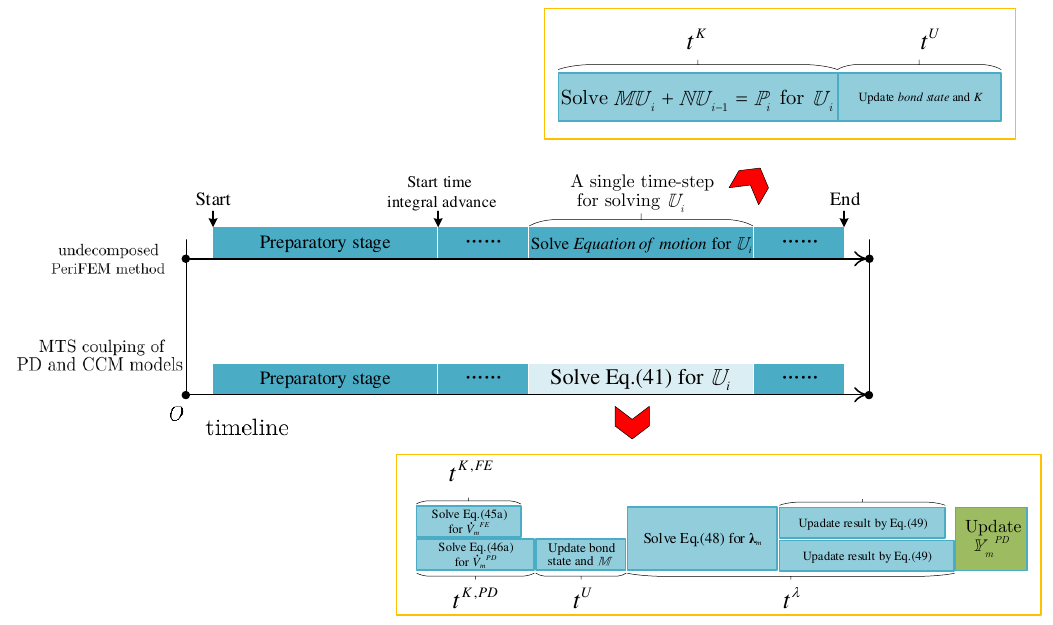}
				\caption[]{The computational time of each part in the solving process.}
				\label{fig.8} 
			\end{center}
		\end{figure}

		In pure PD computation, the solving process is mainly divided into two parts in a single time step. The first step is to solve the dynamic equation to obtain the kinematic quantities $\mathbb{U}_i$ at time $t_i$, and the second step is to update the bond state and global stiffness for the next time step integration in Fig.\ref{fig.8}. The proposed MTS coupling method splits the result on the time scale. Therefore, in a single time step, it is mainly divided into solving the kinematic quantities of two subdomains, solving the Lagrange multiplier force, and finally updating the results. In fact, according to Algorithm 1, when solving the unknowns of two subdomains and updating the results, the two subdomains are independent of each other, meaning they can be solved in parallel. Thus, when solving the CCM and PD models, we record the time consumed by this step as $t^K=max\{t^{K,FE},t^{K,PD}\}$, as shown in Fig.\ref{fig.8}. $\mathbb{Y}$ represents the response of the structure only under a unit load; the result of $\mathbb{Y}$ needs to be updated only when the structure is damaged, and this step can be precalculated using other technologies, for example, machine learning, which will be one of our future work directions, so the time to update $\mathbb{Y}$ is not considered here.
		
		To compare the computational efficiency, we focus on the time consumption in a single time step that solves the motion equations and updates states. As shown in Fig.\ref{fig.8}, $t^K$ represents the time to solve the kinematic quantity equation, and $t^U$ represents the time to update the bond state and global stiffness. In the MTS coupling method, $t^\lambda$ indicates the time to solve the Lagrange multiplier force $\bm{\lambda}$ and update the subdomain results.
		
		The information for the CPU hardware configuration is Intel Core$^{TM}$ i7-12700F with 64G RAM and a single GPU card, which is the Nvidia RTX 3070 with 8GB VRAM. A CUDA C prototype code that achieved the MTS-PDCCM was used for the numerical results.
	
\begin{figure}[H]
			\begin{center}
				\includegraphics[scale=0.9]{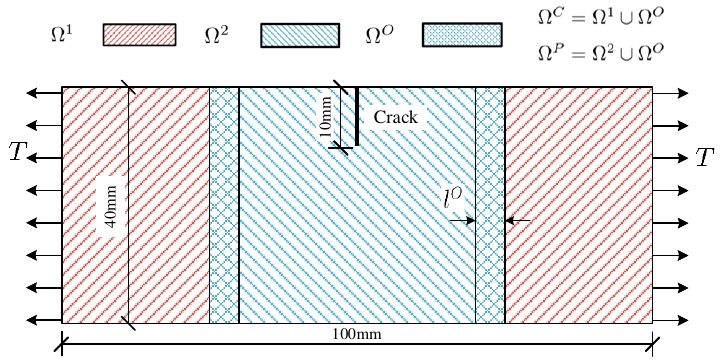}
				\caption[]{The geometry and domain decomposition of the fracture problem.}
				\label{fig.9} 
			\end{center}
		\end{figure}

		We study a mode-I fracture problem of a two-dimensional plate under tension on both sides, as shown in Fig.\ref{fig.9}. The length of the plate is 100 mm, the height is 40 mm, and there is a pre-notch of 10 mm at the upper midpoint. The left and right sides are subjected to a uniform tensile force of $T=16$ MPa. For a pure PD example, it is regarded as a constant body force density along a region of width $\delta$. The properties of the material are $E=72$ GPa, $\nu =1/3$, $\rho=2235$ kg/m$^3$, and $G_0=204$ J/m$^2$. The micromodulus function of this example is still chosen as the exponential form, which is assumed for the plane stress problem as follows:
	
		\begin{equation}
			c(|\bm{\xi}|)=c^0e^{-\frac{|\bm{\xi}|}{l}},\quad c^0=\frac{3E}{\pi \int_{0}^{\delta}r^5e^{-\frac{r}{l}}dr},
		\end{equation}
		where $l=\delta/10$.
	
			Specific computation parameters are listed in Table 2. Note that parameter $\gamma$ in the Newmark-$\beta$ method is 1/2 in all examples. The FE subdomain is chosen to be solved implicitly, i.e., $(\gamma^{FE},\beta^{FE})=(1/2,1/4)$, to ensure the stability of the FE subdomain over large time steps. In Table 2, $\triangle x$ represents the mesh size, $\Omega^P/\Omega$ is the area percentage of the PD subdomain to the total structure, and $m$ is the time ratio, that is, $\triangle t^{FE}=m\triangle t^{PD}$. In all cases, the integral time step of the PD subdomain is $\triangle t^{PD} = 5 \times10 ^{-9} $ s, $\gamma^{PD}= 1/2 $. The last two columns in the table indicate the time consumed in a large time step $\triangle t^{FE}$ and a small time step $\triangle t^{PD}$.
		
			First is the result analysis of cases 1 to 5. Case 1 is a pure PD computation that takes $69.8$ ms to complete a time-step process; Case 2 is a decomposition problem where the area of the PD domain is 20\% of the total area. It is about 23\% faster than the computation of pure PD with $53.9$ ms in a large time step. For cases 3 and 4, we compare the effect of the PD subdomain size on computational efficiency with the time ratio $m=2$. Owing to the area reduction of the PD subdomain, we found that computational efficiency improved in both large and small time steps. Regarding the effect of $m$ on computational efficiency, by comparing the small time steps of cases 4 and 5, we found that $t^{U}/m$ is the same because the PD subdomain size is the same. However, the Lagrange multiplier force $\bm{\lambda}$ only needs to be solved once for each large time step, so as $m$ increases, the time for $t^{\lambda}/m$ decreases. Finally, we also compute and compare the simulations on a smaller grid size $\triangle x=2.5\times 10^{-4}$ in cases 6 and 7. The computational time of the MTS coupling method is about 20\% of that of pure PD in a small time step, that is, 50.6 ms vs. 253.0 ms. A comparison of the computational time of the entire computation shows a difference of 477 s vs. 1500 s, indicating time savings of more than 50\% and significantly improved computational efficiency.
			 
	\begin{table}[t]
		\renewcommand\arraystretch{1.0}
		\begin{center}
			\caption{Computation parameters and time for different cases }
			\begin{tabular}{c|c|c|c|c c |c c }
				\hline
				&case&  $\beta^{PD}$&$\dfrac{\triangle x}{5\times 10^{-4}}$&$\Omega^P/\Omega$&  m & \makecell[c]{Per-time step$\triangle t^{FE}$\\ $t(t^K,t^U,t^{\lambda})({\rm ms})$}&\makecell[c]{Per-time step$\triangle t^{PD}$\\$t(t^U/m,t^{\lambda}/m)({\rm ms})$}\\
				\hline
				Pure PeriFEM&1&$\beta =1/4$&1&\multicolumn{2}{c}{$\triangle t=5\times 10^{-9}$}\vline &\multicolumn{2}{c}{$t(t^K,t^U)=$69.8(10.0,59.8)}\\
				MTS-PDCCM&2			&1/4	&1&20\%				&1		&53.9(4.0,38.4,11.5)   &  49.9(38.4,11.5)\\
					 	 &3			&0		&1&20\%				&2		&95.9(6.6,74.7,14.6)   &  44.7(37.4,7.3)\\
						 &4			&0		&1&10\%				&2		&86.9(5.1,68.9,12.9)   &  41.0(34.5,6.5)\\
						 &5			&0		&1&10\%				&5		&208.7(12.8,174.8,21.1)&  39.2(35.0,4.2)\\
				\hline
				Pure PeriFEM&6&$\beta =0$&1/2&\multicolumn{2}{c}{$\triangle t=5\times 10^{-9}$}\vline &\multicolumn{2}{c}{$t(t^K,t^U)=$253.0(31.3,170.2)}\\
				MTS-PDCCM&7			&0	&1/2&10\%				&10		&552.9(47.2,433.5,72.2)&50.6(43.4,7.2)\\
				\hline
			\end{tabular}
		\end{center}
	\end{table}

\begin{figure}[H]
		\begin{center}
			\includegraphics[scale=0.85]{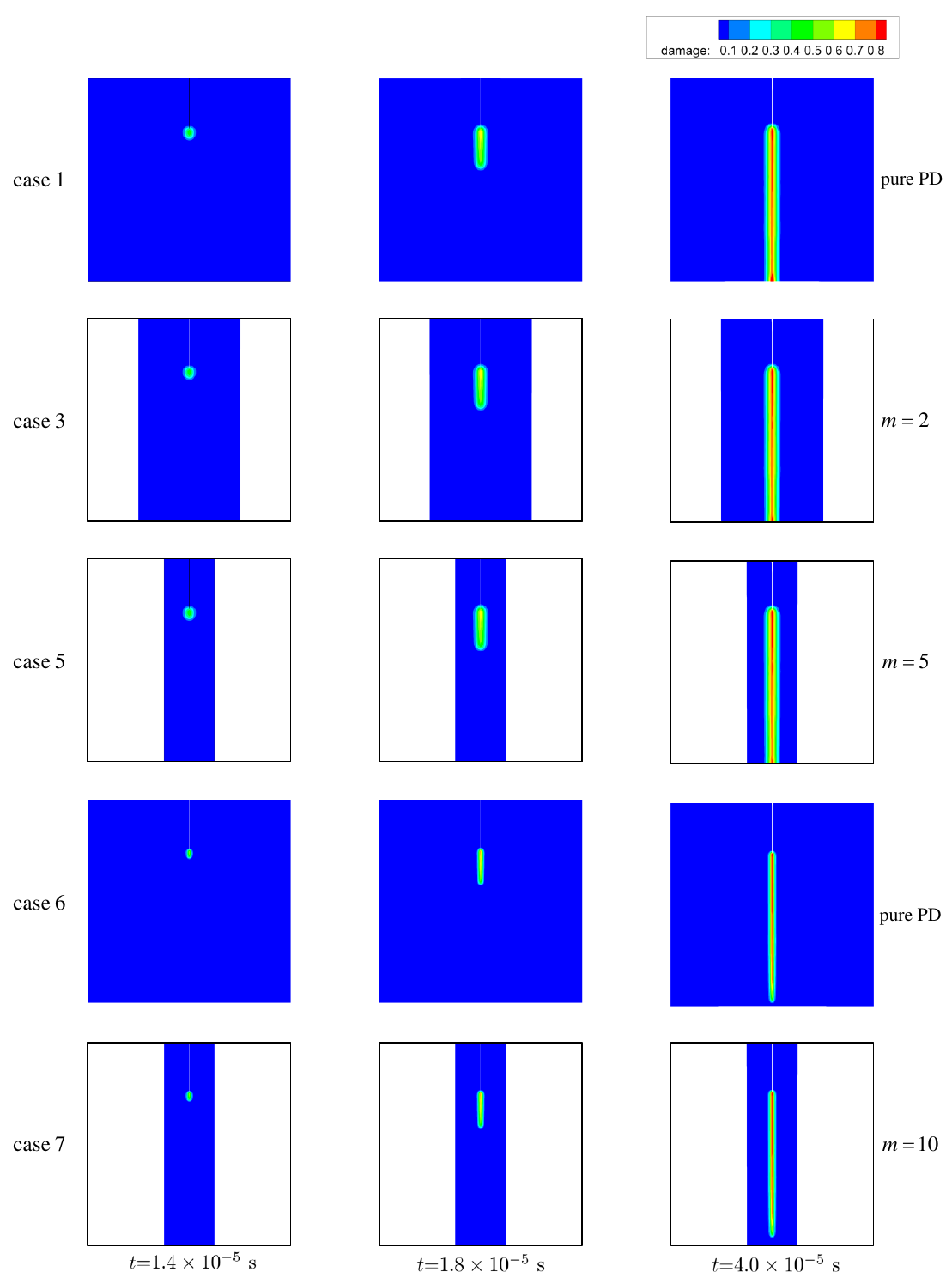}
			\caption[]{The results of the damage for the different cases in Table 2.}
			\label{fig.10} 
		\end{center}
	\end{figure}

	Fig.\ref{fig.10} shows the damage contours of cases 1, 3, 5, 6, and 7 at three times ($t$=$1.4\times 10^{-5}$ s, $1.8\times 10^{-5}$ s, and $4.0\times 10^{-5}$ s). From cases 1, 3, and 5, the length of cracks in each time duration is consistent, indicating that this method can ensure the correctness of structural damage simulation. In addition, the correct results can be obtained for different domain decompositions. As such, the PD subdomain is only limited to the dangerous domain of the structure for larger-scale problems to reduce the unnecessary waste of computing resources.
	Finally, the result comparison of cases 6 and 7 is in explicit simulation; when the time ratio $m$ is larger ($m=10$), the correct result can still be obtained.

	\section{Numerical example}
	In this section, several examples where the PD subdomain covers the cracked area are described.
	\subsection{Crack branching example}
\begin{figure}[H]
		\begin{center}
			\includegraphics[scale=0.8]{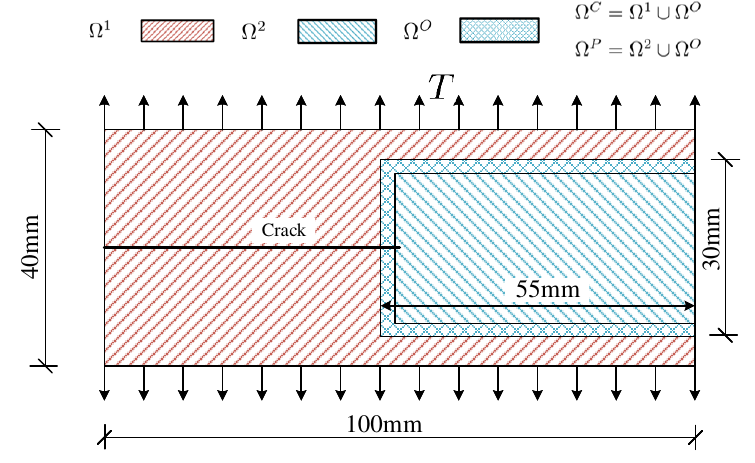}
			\caption[]{Crack branching example: geometry, load conditions, and domain decomposition.}
			\label{fig.11} 
		\end{center}
	\end{figure}

	In the first example, we consider a rectangular plate with a $50$ mm pre-notch whose size is $100\times 40$ mm$^2$, as shown in Fig.\ref{fig.11}. The brittle material used is Duran Glass with mechanical properties: $E = 65$ GPa, $\rho = 2235$ kg/m$^3$, and critical value $s_{crit}=0.002689$. Traction loading $T=12$ MPa is applied to the upper and lower edges during the entire computation. The same material parameters have been used for the pure PD simulation, and the loading is regarded as a constant body force density along a region of width $\delta$. The plate is discretized into 400 × 160 quadrilateral elements; the mesh size is $\triangle x=\triangle y=0.25$ mm. For dynamic parameters, the time steps of the pure PD model are $\triangle t=1\times10^{-8}$ s and $(\gamma,\beta)=(1/2,0)$. For the proposed MTS coupled method, the area of the PD subdomain is $41.25\%$ of the total area with size $55\times 30$ mm$^2$, as shown in Fig.\ref{fig.11}. The width of the overlapping domain is 0.75 mm. The time step of the CCM model is $\triangle t^{FE}=1\times10^{-7}$ s and the time step of the PD model is $\triangle t^{PD}=1\times10^{-8}$ s with time ratio $m=10$, the Newmark-$\beta$ are $(\gamma^{FE},\beta^{FE})=(1/2,1/4)$ and $(\gamma^{PD},\beta^{PD})=(1/2,0)$, respectively. Both examples run for the same total time of $6\times 10^{-5}$ s.

\begin{figure}[H]
	\begin{center}
		\includegraphics[scale=0.85]{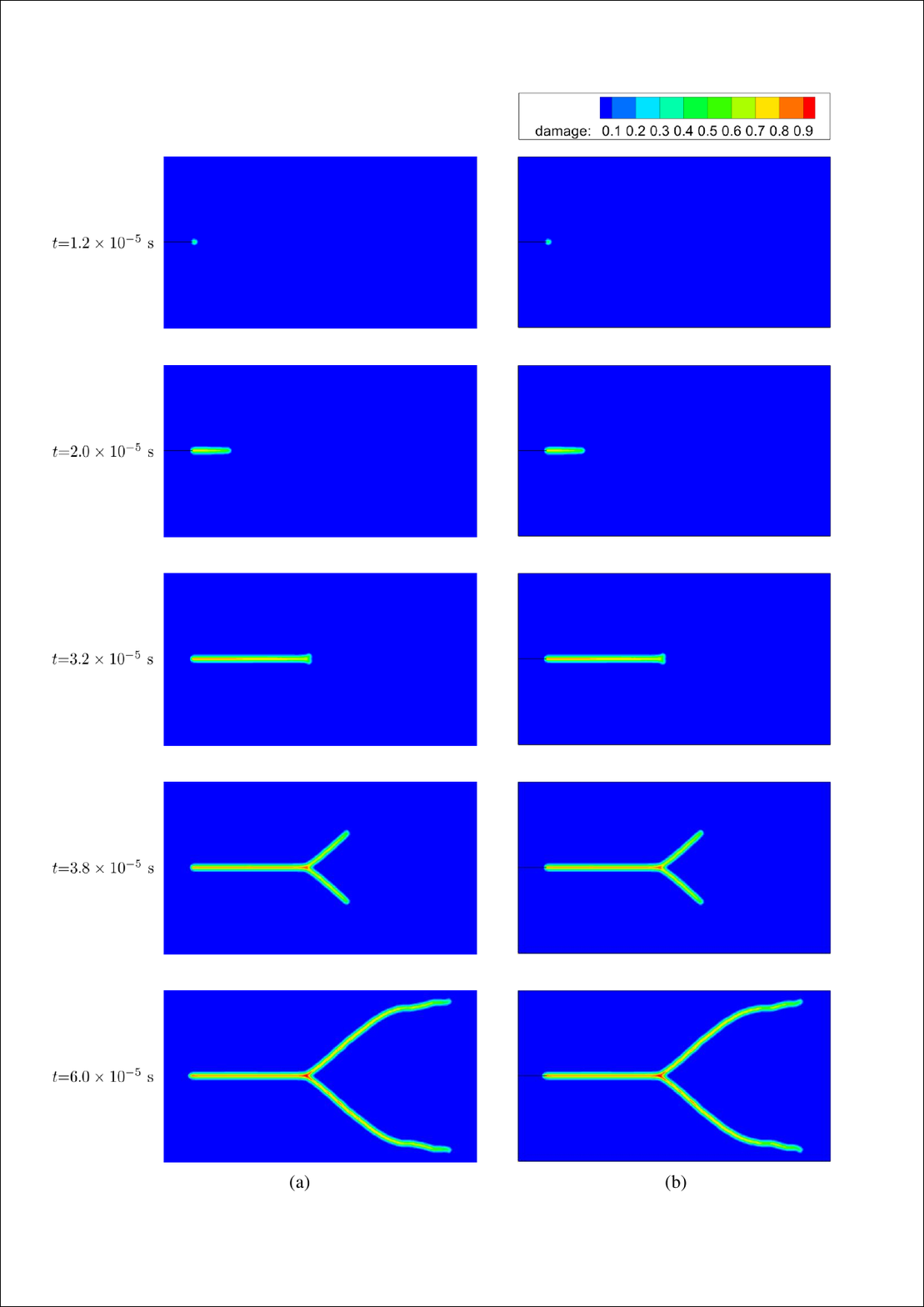}
		\caption[]{Simulated crack path at different times by (a) a pure PD model and (b) the MTS-PDCCM model.}
		\label{fig.12} 
	\end{center}
	\end{figure}

	In Fig.\ref{fig.12} the crack path obtained by the proposed model is compared with that obtained by a pure PD model. At different moments, the crack path is consistent. In terms of computational time consumption, the pure PD model consumes 1320 s, while the proposed model takes only 863 s.
	
	\subsection{Kalthoff–Winkler plate example}
\begin{figure}[H]
		\begin{center}
			\includegraphics[scale=0.9]{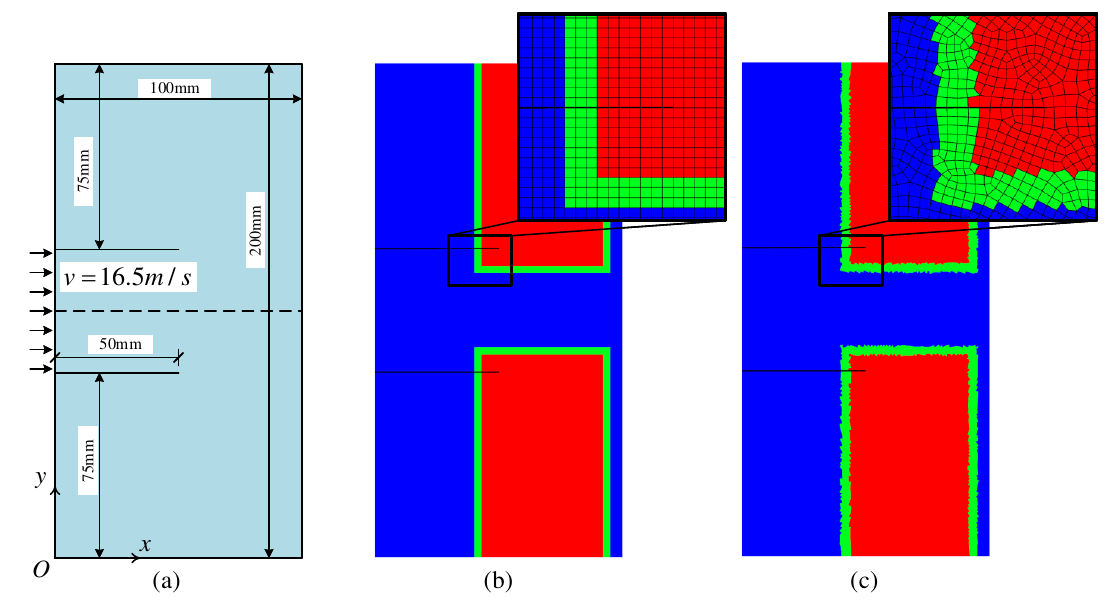}
			\caption[]{Kalthoff–Winkler plate example: (a) geometry and boundary conditions; (b) domain decomposition with structured mesh, (c) domain decomposition with unstructured mesh.}
			\label{fig.13} 
		\end{center}
	\end{figure}

	Another classic example of dynamic fracture is the well-known Kalthoff–Winkler's experiment \cite{KWexper}, which has been successfully simulated by meshfree PD \cite{SILLING2003641}. The geometry and boundary conditions are shown in Fig.\ref{fig.13}(a). The plate is $100\times200$ mm$^2$ with two parallel notches and is assigned a constant speed of $16.5$ m/s in the horizontal direction on the boundary between the notches. We choose a X2 NiCoMo 18-9-5 steel material, and the mechanical properties are $E = 191$ GPa, $\rho= 8000$ kg/m$^3$, and $s_{crit}=0.01$. In this example, we choose structured and unstructured grids for simulation, where the size of the structured mesh is $\triangle x=1$ mm, as shown in Fig.\ref{fig.13}(b), and the average size of the unstructured mesh is about $\triangle x=0.91$ mm, as shown in Fig.\ref{fig.13}(c). For the pure PD, the explicit integration algorithm $(\gamma,\beta)=(1/2,0)$ is used, and the time step is $\triangle=1\times10^{-8}$ s. For the MTS-PDCCM model, two PD subdomains are located around the crack tips on the upper and lower right sides of the plate. In the structured mesh in Fig.\ref{fig.13}(b), the PD subdomains are $\Omega^{PD}=\Omega^1\cup\Omega^2$, where $\{\Omega^1|40<x<95,y>115\}$, $\{\Omega^2|40<x<95,y<85\}$, and the width of overlapping domain is $3$ mm. In an unstructured mesh, the parameters can only be approximated to those of a structured mesh owing to the irregularity of the mesh. For both meshes, the time step of the PD model is $\triangle t^{PD}=1\times10^{-8}$ s with time ratio $m=20$, that is, $\triangle t^{FE}=2\times10^{-7}s$, the Newmark-$\beta$ are $(\gamma^{FE},\beta^{FE})=(1/2,1/4)$ and $(\gamma^{PD},\beta^{PD})=(1/2,0)$, respectively. All examples run for the same total time of $1.1\times 10^{-4}$ s.

	\begin{figure}[H]
		\begin{center}
			\includegraphics[scale=0.95]{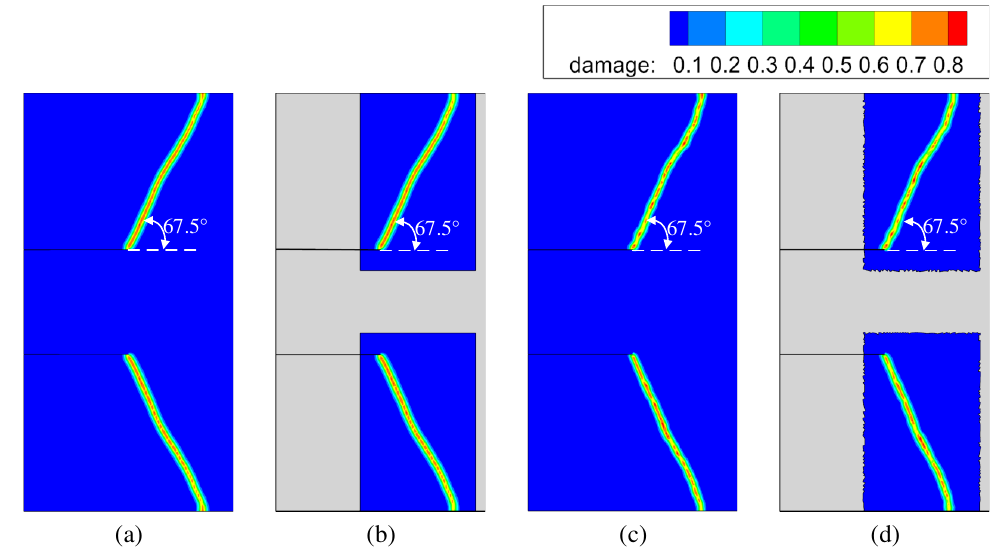}
			\caption[]{The crack path at $t=1.0\times10^{-4}$ s with an inclination angle of $67.5^\circ$(with respect to the x-direction). (a,c) using the pure PD; (b,d) using the MTS-PDCCM model; (a,b) with structured mesh; and (c,d) with unstructured mesh.}
			\label{fig.14} 
		\end{center}
	\end{figure}

	The simulation results are presented in Fig.\ref{fig.14}. We can see that the crack path is the same for a structured grid or an unstructured grid, with an inclined angle of about $67.5^\circ$, which is close to the data from the experimental observation \cite{KWexper}. The entire computational time consumed is 1269 s, 707 s, 1756 s, and 1208 s, as shown in Fig.\ref{fig.14}(a,b,c,and d), respectively.
	
	\subsection{Cracking on a cylinder under internal pressure}
\begin{figure}[H]
		\begin{center}
			\includegraphics[scale=0.9]{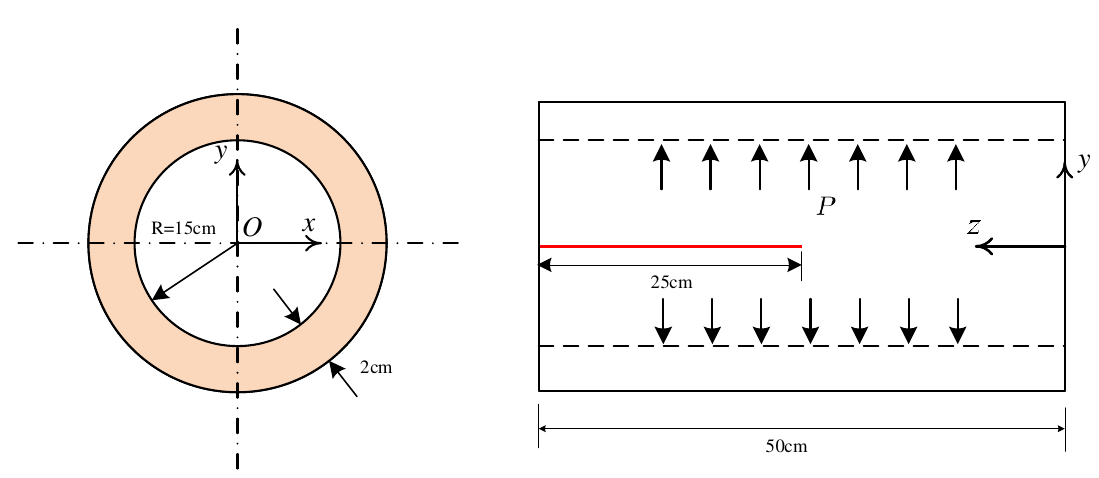}
			\caption[]{Sketch of a 3D cylinder with a crack (red line) under internal pressure.}
			\label{fig.15} 
		\end{center}
	\end{figure}

\begin{figure}[H]
	\begin{center}
		\includegraphics[scale=0.7]{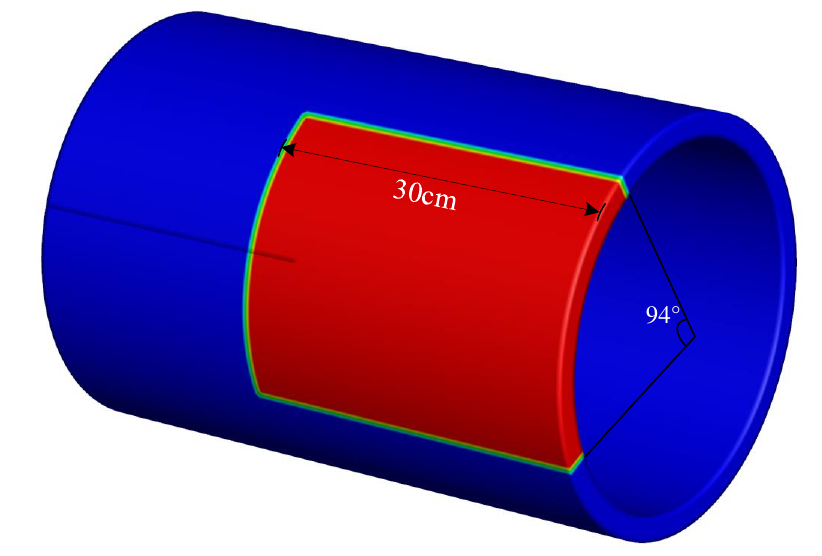}
		\caption[]{Domain decomposition, where blue indicates the FE subdomain, red indicates the PD subdomain, and green indicates the overlapping subdomain.}
		\label{fig.16} 
	\end{center}
	\end{figure}

	The crack growth on a cylinder with a pre-notch under internal pressure, as shown in Fig.\ref{fig.15}, is studied. The cylinder is subjected to a constant internal pressure of $P=9$ MPa. The axial length of the cylinder is $50$ cm, the inner diameter of the section is $15$ cm, and the thickness is $2$ cm. The mechanical properties are $E = 140$ GPa, $\rho = 8000$ kg/m$^3$, $\nu =1/4$, and $s_{crit}=0.004$. For the pure PD model, the horizon $\delta =2.03\triangle x, \triangle x = 0.408$ cm, and the cylinder is discretized with 147500 elements. The explicit integration algorithm $(\gamma,\beta)=(1/2,0)$ is used, and the time step is $\triangle t=2.5\times10^{-7}$ s. For the MTS-PDCCM model, the subdomain decomposition is shown in Fig.\ref{fig.16}, where $\Omega^P$ is $\{\Omega^P|z<0.3, -47^\circ<\theta<47^\circ\}$, $\theta$ is the angle to the opposite X-axis in the xy-plane, and the width of the overlapping domain is $2\triangle x$. The time step of the PD model is $\triangle t^{PD}=2.5\times10^{-7}$ s with time ratio $m=4$, that is, $\triangle t^{FE}=1\times10^{-6}$ s, the Newmark-$\beta$ parameters are $(\gamma^{FE},\beta^{FE})=(1/2,1/4)$ and $(\gamma^{PD},\beta^{PD})=(1/2,0)$, respectively. They both run the same total time of $6\times10^{-4}$ s.

\begin{figure}[H]
	\begin{center}
		\includegraphics[scale=0.75]{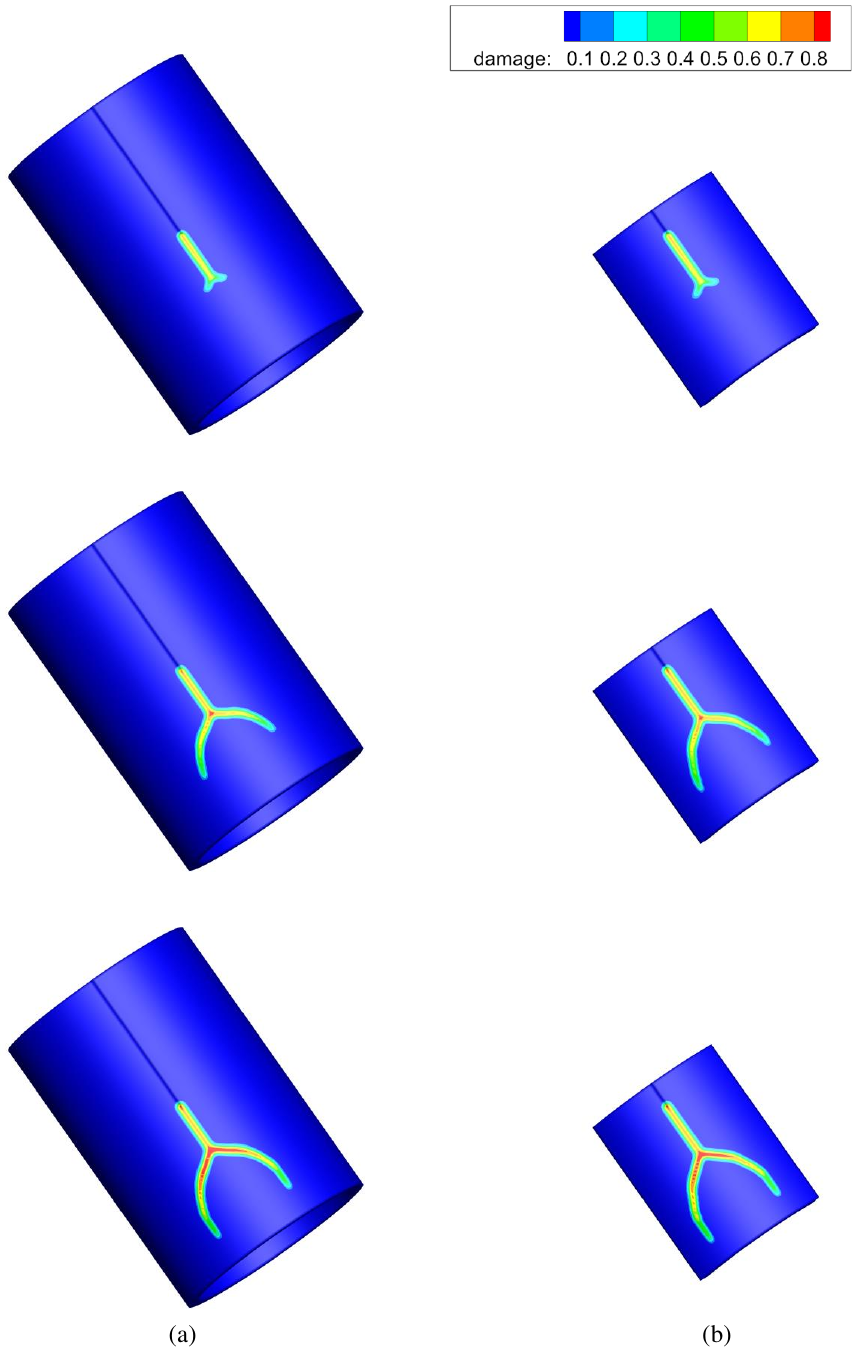}
		\caption[]{The crack paths simulated by (a) the pure PD model and (b) the MTS-PDCCM model. Top row: $t=3\times 10^{-4}$ s; middle row: $t=4.5\times10^{-4}$ s; and bottom row: $t=5.5 \times10^{-4}$ s.}
		\label{fig.17} 
	\end{center}
	\end{figure}

	Several crack paths are shown in Fig.\ref{fig.17}. The crack initially grows along the direction of the pre-notch and then branches. As we can see, the crack results are the same for both the pure PD and MTS-PDCCM models. The pure PD model consumes 9360 s, while the proposed model takes 2710 s.
	\section{Conclusion}
	In this study, the MTS coupling of PD and CCM models is proposed to simulate dynamic fracture, in which the whole computational domain can be decomposed into two or more subdomains, and different subdomains can be computed by different time steps. The PD model is only used in the critical cracking domain, with a small time step. In contrast, the CCM model is used in most other domains and adopts a large time step, significantly reducing the computational cost.
	
	The benchmark examples were performed successfully using this coupling method, demonstrating the validity, high efficiency, and robustness of the proposed coupled model in dynamic brittle fracture. Consequently, this coupling method can be easily applied to large-scale engineering problems. 
	\section*{Acknowledgment}
	The authors gratefully acknowledge the financial support received from the National Natural Science Foundation (12272082), the Foundation for Innovative Research Groups of the National Natural Science Foundation (11821202), and the Fundamental Research Funds for the Central Universities (DUT22QN238)
	\section*{References}
	
	\bibliography{mybibfile}
	
\end{document}